\documentclass[twocolumn]{aastex61}

\usepackage{natbib}
\usepackage{amsmath}	
\usepackage{amssymb}	

\shorttitle{Starburst to quiescent from \textit{HST}/ALMA}
\shortauthors{G\'omez-Guijarro et al.}

\begin{document}

\title{Starburst to quiescent from \textit{HST}/ALMA: \\Stars and dust unveil minor mergers in submillimeter galaxies at $\MakeLowercase{z} \sim 4.5$}

\correspondingauthor{C. G\'omez-Guijarro}
\email{cgguijarro@dark-cosmology.dk}

\author{C. G\'omez-Guijarro}
\affil{Dark Cosmology Centre, Niels Bohr Institute, University of Copenhagen, Juliane Maries Vej 30, DK-2100 Copenhagen, Denmark}
\affil{Cosmic Dawn Center (DAWN), Niels Bohr Institute, University of Copenhagen, Denmark}

\author{S. Toft}
\affil{Dark Cosmology Centre, Niels Bohr Institute, University of Copenhagen, Juliane Maries Vej 30, DK-2100 Copenhagen, Denmark}
\affil{Cosmic Dawn Center (DAWN), Niels Bohr Institute, University of Copenhagen, Denmark}

\author{A. Karim}
\affil{Argelander-Institut f\"ur Astronomie, Universit\"at Bonn, Auf dem H\"ugel 71, D-53121 Bonn, Germany}

\author{B. Magnelli}
\affil{Argelander-Institut f\"ur Astronomie, Universit\"at Bonn, Auf dem H\"ugel 71, D-53121 Bonn, Germany}

\author{G. E. Magdis}
\affil{Dark Cosmology Centre, Niels Bohr Institute, University of Copenhagen, Juliane Maries Vej 30, DK-2100 Copenhagen, Denmark}
\affil{Cosmic Dawn Center (DAWN), Niels Bohr Institute, University of Copenhagen, Denmark}
\affil{Institute for Astronomy, Astrophysics, Space Applications and Remote Sensing, National Observatory of Athens, GR-15236 Athens, Greece}

\author{E. F. Jim\'enez-Andrade}
\affil{Argelander-Institut f\"ur Astronomie, Universit\"at Bonn, Auf dem H\"ugel 71, D-53121 Bonn, Germany}
\affil{International Max Planck Research School of Astronomy and Astrophysics at the Universities of Bonn and Cologne}

\author{P. L. Capak}
\affil{IPAC, 1200 E. California Blvd., Pasadena, CA, 91125, USA}
\affil{California Institute of Technology, 1200 E. California Blvd., Pasadena, CA 91125, USA}

\author{F. Fraternali}
\affil{Department of Physics and Astronomy, University of Bologna, viale Bertu Pichat 6/2, I-40127 Bologna, Italy}
\affil{Kapteyn Astronomical Institute, University of Groningen, P.O. Box 800, 9700AV Groningen, The Netherlands}

\author{S. Fujimoto}
\affil{Institute for Cosmic Ray Research, The University of Tokyo, Kashiwa, Chiba 277-8582, Japan}

\author{D. A. Riechers}
\affil{Department of Astronomy, Cornell University, Space Science Building, Ithaca, NY 14853, USA}

\author{E. Schinnerer}
\affil{Max Planck Institut f\"ur Astronomie, K\"onigtuhl 17, D-69117, Heidelberg, Germany}

\author{V. Smol\v{c}i\'c}
\affil{Department of Physics, Faculty of Science, University of Zagreb,  Bijeni\v{c}ka cesta 32, 10000  Zagreb, Croatia}

\author{M. Aravena}
\affil{N\'ucleo de Astronom\'ia, Facultad de Ingenier\'ia y Ciencias, Universidad Diego Portales, Av. Ej\'ercito 441, Santiago, Chile}

\author{F. Bertoldi}
\affil{Argelander-Institut f\"ur Astronomie, Universit\"at Bonn, Auf dem H\"ugel 71, D-53121 Bonn, Germany}

\author{I. Cortzen}
\affil{Dark Cosmology Centre, Niels Bohr Institute, University of Copenhagen, Juliane Maries Vej 30, DK-2100 Copenhagen, Denmark}
\affil{Cosmic Dawn Center (DAWN), Niels Bohr Institute, University of Copenhagen, Denmark}

\author{G. Hasinger}
\affil{Institute for Astronomy, University of Hawaii, 2680 Woodlawn Dr, Honolulu, HI 96822, USA}

\author{E. M. Hu}
\affil{Institute for Astronomy, University of Hawaii, 2680 Woodlawn Dr, Honolulu, HI 96822, USA}

\author{G. C. Jones}
\affil{Physics Department, New Mexico Institute of Mining and Technology, Socorro, NM 87801, USA}
\affil{National Radio Astronomy Observatory, 1003 Lopezville Road, Socorro, NM 87801, USA}

\author{A. M. Koekemoer}
\affil{Space Telescope Science Institute, 3700 San Martin Drive, Baltimore, MD 21218, USA}

\author{N. Lee}
\affil{Dark Cosmology Centre, Niels Bohr Institute, University of Copenhagen, Juliane Maries Vej 30, DK-2100 Copenhagen, Denmark}

\author{H. J. McCracken}
\affil{Sorbonne Universit\'es, UPMC Univ Paris 06, UMR 7095, Institut d'Astrophysique de Paris F-75014 Paris, France}
\affil{CNRS, UMR 7095, Institut d'Astrophysique de Paris, F-75014 Paris, France}

\author{M. J. Micha\l{}owski}
\affil{Astronomical Observatory Institute, Faculty of Physics, Adam Mickiewicz University, ul. S\l{}oneczna 36, 60-286 Pozna\'n, Poland}

\author{F. Navarrete}
\affil{Max Planck Institut f\"ur Radioastronomie, Auf dem H\"ugel 69, D-53121 Bonn, Germany}

\author{M. Povi\'c}
\affil{Ethiopian Space Science and Technology Institute (ESSTI), Entoto Observatory and Research Center (EORC), Astronomy and Astrophysics Research Division, P.O. Box 33679, Addis Ababa, Ethiopia}
\affil{Instituto de Astrof\'isica de Andaluc\'ia (IAA-CSIC), Glorieta de la Astronom\'ia s/n, E-18008 Granada, Spain}

\author{A. Puglisi}
\affil{European Southern Observatory, Karl-Schwarzschild-Strasse 2, D-85748 Garching bei M\"unchen, Germany}

\author{E. Romano-D\'iaz}
\affil{Argelander-Institut f\"ur Astronomie, Universit\"at Bonn, Auf dem H\"ugel 71, D-53121 Bonn, Germany}

\author{K. Sheth}
\affil{NASA Headquarters, 300 E Street SW, Washington DC 20546, USA}

\author{J. D. Silverman}
\affil{Kavli Institute for the Physics and Mathematics of the Universe (WPI), The University of Tokyo Institutes for Advanced Study, The University of Tokyo, Kashiwa, Chiba 277-8583, Japan}

\author{J. Staguhn}
\affil{NASA Goddard Space Flight Center, Code 665, Greenbelt, MD 20771, USA}
\affil{Department of Physics and Astronomy, Johns Hopkins University, Baltimore, MD 21218, USA}

\author{C. L. Steinhardt}
\affil{Dark Cosmology Centre, Niels Bohr Institute, University of Copenhagen, Juliane Maries Vej 30, DK-2100 Copenhagen, Denmark}
\affil{Cosmic Dawn Center (DAWN), Niels Bohr Institute, University of Copenhagen, Denmark}

\author{M. Stockmann}
\affil{Dark Cosmology Centre, Niels Bohr Institute, University of Copenhagen, Juliane Maries Vej 30, DK-2100 Copenhagen, Denmark}
\affil{Cosmic Dawn Center (DAWN), Niels Bohr Institute, University of Copenhagen, Denmark}

\author{M. Tanaka}
\affil{National Astronomical Observatory of Japan, 2-21-1 Osawa, Mitaka, Tokyo 181-8588, Japan}

\author{F. Valentino}
\affil{Dark Cosmology Centre, Niels Bohr Institute, University of Copenhagen, Juliane Maries Vej 30, DK-2100 Copenhagen, Denmark}
\affil{Cosmic Dawn Center (DAWN), Niels Bohr Institute, University of Copenhagen, Denmark}

\author{E. van Kampen}
\affil{European Southern Observatory, Karl-Schwarzschild-Strasse 2, D-85748 Garching bei M\"unchen, Germany}

\author{A. Zirm}
\affil{Dark Cosmology Centre, Niels Bohr Institute, University of Copenhagen, Juliane Maries Vej 30, DK-2100 Copenhagen, Denmark}
\affil{Greenhouse Software, 3rd Floor, 110 5th Avenue, New York, NY 10011, USA}

\begin{abstract}

Dust-enshrouded, starbursting, submillimeter galaxies (SMGs) at $z \geq 3$ have been proposed as progenitors of $z \geq 2$ compact quiescent galaxies (cQGs). To test this connection, we present a detailed spatially resolved study of the stars, dust and stellar mass in a sample of six submillimeter-bright starburst galaxies at $z \sim 4.5$. The stellar UV emission probed by \textit{HST} is extended, irregular and shows evidence of multiple components. Informed by \textit{HST}, we deblend \textit{Spitzer}/IRAC data at rest-frame optical finding that the systems are undergoing minor mergers, with a typical stellar mass ratio of 1:6.5. The FIR dust continuum emission traced by ALMA locates the bulk of star formation in extremely compact regions (median $r_{\rm{e}} = 0.70 \pm 0.29$\,kpc) and it is in all cases associated with the most massive component of the mergers (median $\log (M_{*}/M_{\odot}) = 10.49 \pm 0.32$). We compare spatially resolved UV slope ($\beta$) maps with the FIR dust continuum to study the infrared excess ($\rm{IRX} = L_{\rm{IR}}/L_{\rm{UV}}$)-$\beta$ relation. The SMGs display systematically higher $\rm{IRX}$ values than expected from the nominal trend, demonstrating that the FIR and UV emissions are spatially disconnected. Finally, we show that the SMGs fall on the mass-size plane at smaller stellar masses and sizes than cQGs at $z = 2$. Taking into account the expected evolution in stellar mass and size between $z = 4.5$ and $z = 2$ due to the ongoing starburst and mergers with minor companions, this is in agreement with a direct evolutionary connection between the two populations.

\end{abstract}

\keywords{galaxies: evolution --- galaxies: formation --- galaxies: high-redshift --- galaxies: interactions --- galaxies: ISM --- galaxies: starburst --- galaxies: structure --- infrared: galaxies --- submillimeter: galaxies --- ultraviolet: galaxies}

\section{Introduction} \label{sec:intro}

Giant elliptical galaxies are the oldest, most massive galaxies in the local Universe. Understanding their formation and evolution is one of the major challenges in contemporary galaxy evolution studies. They are uniformly old, red and quiescent, i.e., void of star formation. Studies of their stellar populations suggests that they formed in violent bursts of star formation at $z \sim 3$--5 \citep[e.g.,][]{2005ApJ...621..673T}. Their evolution has been traced all the way back to $z \sim 4$ through the study of mass complete samples of quiescent galaxies as a function of redshift \citep[e.g.,][]{2011ApJ...739...24B,2014ApJ...788...28V,2015ApJ...808L..29S,2017A&A...605A..70D}.

Compared with their lower redshift descendants, at $z \sim 2$ half of the most massive galaxies are already old, quiescent and are furthermore found to be extremely compact systems \citep[e.g.,][]{2007ApJ...671..285T,2008ApJ...677L...5V,2012ApJ...749..121S}. The brightest examples of these compact quiescent galaxies (cQGs) at $z \sim 2$ (for which follow-up spectroscopy has been possible) show clear post-starburst features, evidence of a starburst at $z > 3$ \citep[e.g.,][Stockmann et al., in prep]{2012ApJ...754....3T,2013ApJ...771...85V,2016Natur.540..248K,2017ApJ...834...18B,2017Natur.546..510T}. Their subsequent evolution into local ellipticals is most likely dominated by passive aging of their stellar populations and merging with minor companions \citep[e.g.,][]{2009ApJ...697.1290B,2012ApJ...744...63O,2012ApJ...746..162N,2012ApJ...754....3T}.

The most intense starbursts known are the so-called dusty star-forming galaxies (DSFGs), which are characterized by star formation rates of up to thousands of solar masses per year \citep[see][for a review]{2014PhR...541...45C}. The best studied DSFGs are the submillimeter galaxies (SMGs) \citep[e.g.,][]{2002PhR...369..111B}. Their high dust content absorbs the intense ultraviolet (UV) emission from the starburst and re-radiates it at far-infrared/submillimeter (FIR/sub-mm) wavelengths \citep[e.g.,][]{2014MNRAS.438.1267S}, making the most intense starbursts easily detectable in sub-mm surveys to the highest redshift.

Following the discovery of a high-redshift tail in the SMGs redshift distribution \citep[e.g.,][]{2005ApJ...622..772C,2008ApJ...681L..53C,2011Natur.470..233C,2009ApJ...694.1517D,2012A&A...548A...4S,2013ApJ...767...88W,2015A&A...577A..29M,2016ApJ...822...80S,2017A&A...608A..15B}, \citet{2014ApJ...782...68T} presented evidence for a direct evolutionary connection between $z \gtrsim 3$ SMGs and $z \sim 2$ cQGs based on the formation redshift distribution for the quiescent galaxies, number density arguments and the similarity of the distributions of the two populations in the stellar mass-size plane \citep[see also e.g.,][]{2008A&A...482...21C,2014ApJ...788..125S,2015ApJ...799...81S,2015ApJ...810..133I,2016ApJ...833..103H,2016ApJ...827...34O,2017arXiv170904191O}. However, as the latter was based on sizes derived from low resolution data probing the rest-frame UV emission (which is likely biased towards unobscured, young stellar populations), confirmation using higher quality data is crucial.

To test the proposed evolutionary connection, we here present deep, high resolution \textit{Hubble Space Telescope} (\textit{HST}) and Atacama Large Millimeter/submillimeter Array (ALMA) follow-up observations of six of the highest-redshift SMGs from \citet{2014ApJ...782...68T}, five of which are spectroscopically confirmed at $z \sim 4.5$. The data probe the distribution of the UV-bright stellar populations and the FIR dust continuum emission, which allows for a full characterization of the star formation and dust attenuation in the galaxies. The sources are drawn from the COSMOS field, thus a wealth of deep ground- and space-based lower resolution optical--mid-IR data are available, which we use to obtain stellar masses for the systems.

In two companion papers we will explore the gas/dust distributions and kinematics of the sample (Karim et al., in prep) and the detailed molecular gas properties of one of the sources \citep{2017arXiv171010181J}.

The layout of the paper is as follows. We introduce the sample, data and methodology in Section~\ref{sec:sample_data}. In Section~\ref{sec:morph} we present the rest-frame UV/FIR morphologies of the sample. The results based on the comparison of the dust as seen in absorption and emission are shown in Section~\ref{sec:dust}. Stellar masses are discussed in Section~\ref{sec:mstar}. We show the evolutionary connection between SMGs and cQGs in Section~\ref{sec:mass_size}. Additional discussion is presented in Section~\ref{sec:discussion}. We summarize the main findings and conclusions in Section~\ref{sec:summary}.

Throughout this work we adopted a concordance cosmology $[\Omega_\Lambda,\Omega_M,h]=[0.7,0.3,0.7]$ and Chabrier initial mass function (IMF) \citep{2003PASP..115..763C}. The AB magnitude system was employed across the whole study \citep{1974ApJS...27...21O}.

\section{Sample and Data} \label{sec:sample_data}

\subsection{COSMOS SMGs Sample} \label{subsec:sample}

We selected a sample of six of the highest-redshift unlensed SMGs from \citet{2014ApJ...782...68T} (see Table~\ref{tab:sample}), which are part of the extensive (sub)millimeter interferometric and optical/millimeter spectroscopic follow-up campaings in the COSMOS field \citep{2007ApJS..172....1S,2007ApJ...671.1531Y,2008ApJ...688...59Y,2008ApJ...681L..53C,2011Natur.470..233C,2008ApJ...689L...5S,2010ApJ...720L.131R,2014ApJ...796...84R,2011ApJ...731L..27S,smolcic15,2015MNRAS.454.3485Y}. All our sample sources had been spectroscopically confirmed to be at $4.3 \lesssim z \lesssim 4.8$, except AzTEC5 at a slightly lower (photometric) redshift (see Table~\ref{tab:prop}). We refer the reader to \citet{smolcic15} for a detailed description of the selection of each source.

\begin{deluxetable}{lccc}
\tabletypesize{\scriptsize}
\tablecaption{Sample of Targetted SMGs in COSMOS. \label{tab:sample}}
\tablehead{\colhead{Source Name} & \colhead{Other Name} & \colhead{$\alpha$(J2000)\tablenotemark{a}} & \colhead{$\delta$(J2000)\tablenotemark{a}} \\ \colhead{} & \colhead{} & \colhead{[h:m:s]} & \colhead{[$^\circ$:$'$:$''$]}}
\startdata 
AK03       &        & 10 00 18.74 & +02 28 13.53 \\
AzTEC1     & AzTEC/C5       & 09 59 42.86 & +02 29 38.2 \\
AzTEC5     & AzTEC/C42       & 10 00 19.75 & +02 32 04.4 \\
AzTEC/C159 &        & 09 59 30.42 & +01 55 27.85 \\
J1000+0234 & AzTEC/C17       & 10 00 54.48 & +02 34 35.73 \\
Vd-17871   &        & 10 01 27.08 & +02 08 55.60 \\
\enddata
\tablenotetext{a}{From \citet{smolcic17}: AK03, AzTEC/C159 and Vd-17871 refer to the VLA 3\,GHz peak position \citep{smolcic15}; AzTEC1 and AzTEC5 refer to the SMA 890\,$\mu$m peak position \citep{2007ApJ...671.1531Y}; J1000+0234 refer to the PdBI $^{12}$CO(4-3) emission line peak position \citep{2008ApJ...689L...5S}.}
\end{deluxetable}

\subsection{\textit{HST} Data} \label{subsec:hst_data}

\textit{HST} WFC3/IR observations of AzTEC1, J1000+0234 and Vd-17871 were taken in the $F125W$ and $F160W$ bands at a 2-orbit depth on each filter (program 13294; PI: A. Karim). For AK03 and AzTEC5, WFC3 $F125W$ and $F160W$ imaging were taken from the CANDELS survey \citep{2011ApJS..197...35G,2011ApJS..197...36K}. AzTEC/C159 was not in our \textit{HST} program due to its faintness at near-IR wavelengths. Additionally, we included  COSMOS \textit{HST} ACS/WFC $F814W$ images \citep{2007ApJS..172..196K} available for the full sample. At the redshift of the sources, these three bands probe the UV continuum regime in the range $\sim 140$--300\,nm (175--345\,nm for AzTEC5).

In order to process the \textit{HST} observations from our program we made use of the DrizzlePac 2.0 package \citep{2012drzp.book.....G}. First, we assured a good alignment between the four dithered frames on each band using the \texttt{TweakReg} task. Next, we combined the frames with \texttt{AstroDrizzle} employing the same parameters as used in the CANDELS reduction procedure: $\rm{\texttt{final\_scale}} = 0.06$ and $\rm{\texttt{final\_pixfrac}} = 0.8$ \citep{2011ApJS..197...36K}.

For the purpose of this work, it is important that all three bands are properly aligned sharing a common World Coordinate System (WCS) frame with accurate absolute astrometry. In order to guarantee the absolute astrometric accuracy we chose the COSMOS ACS $F814W$ image as the reference frame. The fundamental astrometric frame for COSMOS uses the CFHT Megacam $i$-band image \citep{2007ApJS..172...99C}. The latter is tied to the USNO-B1.0 system \citep{2003AJ....125..984M}, which is also tied to the VLA 1.4\,GHz image \citep{2004AJ....128.1974S}, ensuring an absolute astrometric accuracy of 0\farcs05--0\farcs1 or better, corresponding to $\sim 1$--1.5\,pix for our pixel scale. To align the $F125W$ and $F160W$ images to the $F814W$ WCS, we used \texttt{TweakReg} along with SExtractor \citep{1996A&AS..117..393B} catalogs of the three bands, with the $F814W$ catalog and frame as references. Once the three bands shared the same WCS frame, we propagated the WCS solution back to the original \texttt{flt.fits} frames using the \texttt{TweakBack} task, and then ran \texttt{AstroDrizzle} once again to produce the final drizzled images. In the case of AK03 and AzTEC5, where the $F125W$ and $F160W$ data came from CANDELS, this alignment procedure is not necessary since the images are already matched to the COSMOS WCS. The final drizzled images in the three bands were resampled to a common grid and a pixel scale of 0\farcs06\,pix$^{-1}$ using SWarp \citep{2002ASPC..281..228B}.

\subsection{ALMA Data} \label{subsec:alma_data}

Our galaxies were observed in ALMA's Cycle-2 as part of the Cycle-1 (program 2012.1.00978.S; PI: A. Karim). We used the ALMA band-7 and tuned the correlator such that a single spectral window (SpW) would cover the [\ion{C}{2}] line emission of our galaxies, while three adjacent SpWs with a total bandwidth of 5.7\,GHz would be used for continuum detection. These continuum SpWs are those analysed in our study, while the [\ion{C}{2}] line datacubes are presented in Karim et al. (in prep).

Observations were all taken in June 2014, using 34 12-m antennae in configuration C34-4 with a maximum baseline of $\sim 650$\,m. For all galaxies, J1058+0133 and J1008+0621 were used as bandpass and phase calibrators, respectively. In contrast, the flux calibrator is not the same for all galaxies, varying from Titan, J1058+0133, Ceres or Pallas. Calibration was performed with the Common Astronomy Software Applications (CASA; version 4.2.2) using the scripts provided by the ALMA project. Calibrated visibilities were systematically inspected and additional flaggings were added to the original calibration scripts. Flux calibrations were validated by checking the flux density accuracies of our phase and bandpass calibrators. Continuum images were created by combining the three adjacent continuum SpWs with the CASA task \texttt{CLEAN} in multi-frequency synthesis imaging mode and using a standard Briggs weighting scheme with a robust parameter of -1.0. The effective observing frequencies, synthesized beams and resulting noise of these continuum images are listed in Table~\ref{tab:alma}.

\begin{deluxetable*}{lccccc}
\tabletypesize{\scriptsize}
\tablecaption{ALMA Continuum Images Properties. \label{tab:alma}}
\tablehead{\colhead{Source Name} & \colhead{$\nu_{\rm{obs}}$} & \colhead{Beam Size} & \colhead{$\sigma$} & \colhead{$S_{870}^{\rm{ALMA}}$}\tablenotemark{a} & \colhead{$S_{850}^{\rm{SCUBA2}}$}\tablenotemark{b} \\ \colhead{} & \colhead{[Ghz]} & \colhead{[\arcsec$\times$\arcsec]} & \colhead{[mJy\,beam$^{-1}$]} & \colhead{[mJy]} & \colhead{[mJy]}}
\startdata 
AK03       & 337.00 & 0.29$\times$0.27 & 0.17 & 2.3(2.7) $\pm$ 0.2 & 2.4(1.7) $\pm$ 0.6 \\
AzTEC1     & 344.67 & 0.25$\times$0.22 & 0.47 & 14.5(15.7) $\pm$ 0.2 & 14.8(14.3) $\pm$ 1.2 \\
AzTEC5     & 301.78 & 0.47$\times$0.28 & 0.089 & 7.2(12.4) $\pm$ 0.2 & 13.2(13.1) $\pm$ 0.7 \\
AzTEC/C159 & 349.67 & 0.28$\times$0.27 & 0.20 & 6.9(7.1) $\pm$ 0.2 & 6.8(5.5) $\pm$ 1.3 \\
J1000+0234 & 349.85 & 0.30$\times$0.23 & 0.11 & 7.6(7.8) $\pm$ 0.2 & 6.7(5.8) $\pm$ 1.0 \\
Vd-17871   & 345.75 & 0.35$\times$0.31 & 0.21 & 5.2(5.6) $\pm$ 0.2 & 4.8(3.9) $\pm$ 0.9 \\
\enddata
\tablenotetext{a}{In brackets conversion into 850\,$\mu$m fluxes assuming a standard Rayleigh-Jeans slope of 3.5.}
\tablenotetext{b}{In brackets deboosted fluxes.}
\end{deluxetable*}

Each galaxy yields a significant continuum detection $\rm{S/N} > 10$ at the phase center of our images. Their fluxes were measured via 2D Gaussian fits using the python package \texttt{PyBDSF} and are given in Table~\ref{tab:alma}. These fluxes are consistent with those measured 850\,$\mu$m fluxes from the S2COSMOS/SCUBA2 survey (Simpson et al., in prep). This suggests that there is not extended emission which is resolved out in the higher resolution ALMA observations.

In terms of the WCS, we do not expect a significant offset in the ALMA absolute astrometry with respect to the COSMOS WCS. The main source of uncertainty for the relative astrometry between ALMA and \textit{HST} is the uncertainty in the \textit{HST} absolute astrometry with respect to the COSMOS WCS, which is $< 0\farcs1$ as shown in the previous section. \citet{2017arXiv170903505S} tested the relative astrometry between an ALMA single pointing and an \textit{HST} image tied to the COSMOS WCS. Following \citet{2017arXiv170903505S}, at our S/N and resolution the combined pointing accuracy between our ALMA and \textit{HST} images is $< 0\farcs12$, corresponding to $< 2$\,pix for our pixel scale.

\subsection{PSF Matching} \label{subsec:psfmatch}

The \textit{HST} data span three different bands from two different instruments, so consequently, the spatial resolution is different. It is essential to compare the same physical regions when obtaining resolved color information. We therefore degraded the ACS $F814W$ and WFC3 $F125W$ images to the resolution of the WFC3 $F160W$ data (0\farcs18 FWHM), which has the broadest point-spread function (PSF). First, we created a stacked PSF in the different bands, selecting stars that were not saturated and that did not show irregularities on their light profiles. Second, we derived the kernels to match the ACS $F814W$ and WFC3 $F125W$ PSFs to the PSF in the WFC3 $F160W$ image using the task \texttt{PSFMATCH} in IRAF. We applied a cosine bell function tapered in frequecy space to avoid introducing artifacts in the resulting kernel from the highest frequencies. To get the best size for the convolution box we iterated over different values. Finally, we implemented the kernel on the ACS $F814W$ and WFC3 $F125W$ images. The matched PSFs FWHM in the different bands deviate by less than 2\%.

ALMA continuum images also show different spatial resolution compared to that in the PSF-matched \textit{HST} images (median synthesized beam size of 0\farcs30$\times$0\farcs27 versus 0\farcs18 FWHM, respectively). It is important to perform the measurements in the same physical regions when comparing \textit{HST} and ALMA photometry as well, such as to derive rest-frame FIR/UV ratios. When this is required, we used \textit{HST} images matched to the resolution of the ALMA continuum images constructed following the same procedure explained above. In this case the kernel was computed from the WFC3 $F160W$ PSF and the ALMA cleam beam, and then applied to the PSF-matched \textit{HST} images. The matched PSFs FWHM in the \textit{HST} and ALMA images deviate by less than 2\%.

\subsection{Adaptative Smoothing} \label{subsec:smooth}

We applied a smoothing technique to the \textit{HST} images to enhance the signal-to-noise ratio (S/N) and improve our ability to detect low surface brightness features and color gradients between neighboring pixels.

The code employed for this purpose was \texttt{ADAPTSMOOTH} \citep{2009arXiv0911.4956Z}, which smooths the images in an adaptative fashion, meaning that at any pixel only the minimum smoothing length to reach the S/N requested is applied. In this way the images retain the original resolution in regions where the S/N is high and only low S/N regions are smoothed.  

We required a minimum $\rm{S/N} = 5$ and a maximum smoothing length of two neighboring pixels in the code. The former holds true for uncorrelated noise, which is not the case for the drizzled \textit{HST} images analyzed here. In our images the chosen value of 5 corresponds to $\rm{S/N} \sim 2$ when taking into account the noise correlation and pixelation effects in the code. The chosen smoothing length prevents cross-talking between pixels, also reduced by calculating the median of the pixel distribution inside the smoothing radius as opposed to the mean. Such a smoothing length was chosen to match the resolution in the \textit{HST} data, so the smoothing technique does not smear out the images.

We generated a smoothing mask for each band, which is a mask of the required smoothing length to reach the requested minimum S/N for each pixel. When applying a mask to the images, the pixels that do not reach the minimum S/N level are blanked out by the code. If a pixel reached the minimum S/N in at least two bands, we replaced the smoothing length in the mask by the maximum value of them. This guarantees that the same physical regions are probed in different bands, maintaining at the same time the signal if a pixel is above the minimum S/N only in one band.

\subsection{Additional Photometric Data} \label{subsec:add_data}

A series of additional multiwavelength imaging datasets in the optical/IR were employed in this work: Subaru Hyper Suprime-Cam (HSC) from the HSC Subaru Strategic Program (SSP) team and the University of Hawaii (UH) joint dataset in $g$, $r$, $i$, $z$ and $y$ bands \citep{2017arXiv170600566T}, with spatial resolution (seeing FWHM) of 0\farcs92, 0\farcs57, 0\farcs63, 0\farcs64 and 0\farcs81, respectively; the UltraVISTA DR3 survey \citep{2012A&A...544A.156M} covering near-IR $J$, $H$ and $K_{\rm{s}}$ bands, which have resolution of 0\farcs8, 0\farcs7 and 0\farcs7, respectively; and the \textit{Spitzer} Large Area Survey with Hyper-Suprime-Cam (SPLASH; Capak et al., in prep.) mid-IR \textit{Spitzer}/IRAC 3.6 and 4.5\,$\mu$m, with a PSF FWHM of 1\farcs66 and 1\farcs72, respectively.

\section{Morphology} \label{sec:morph}

\begin{figure*}
\centering
\includegraphics[width=\textwidth]{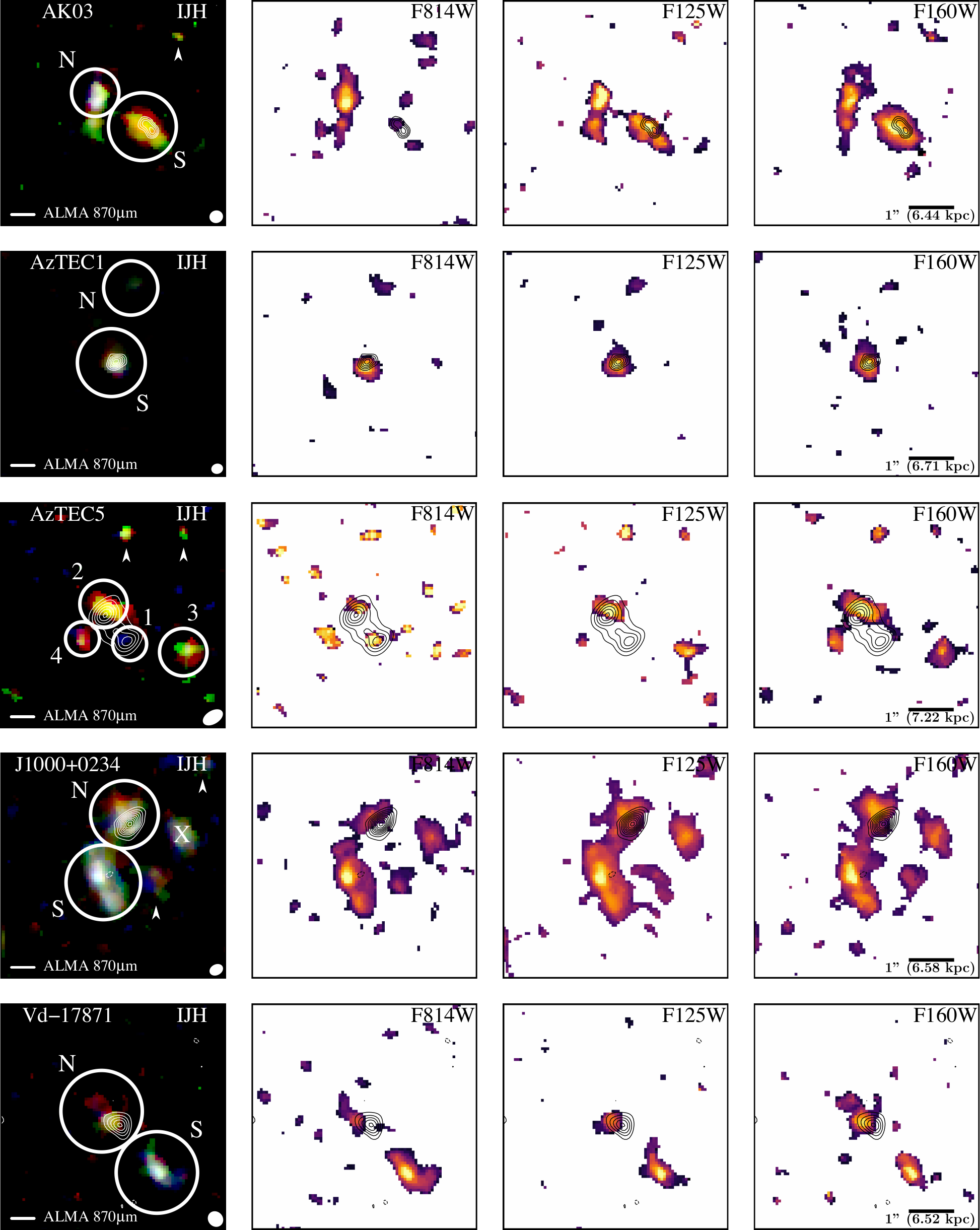}
\caption{$F814W$, $F125W$ and $F160W$ images of the five sources observed with \textit{HST}, and RGB color composite assembled from these three bands. ALMA band 7 ($\sim 870$\,$\mu$m) contours are overlayed. The images are scaled from $\rm{S/N} = 2$ to 75\% of the peak value. The contours shown start at $\pm3\sigma$ and go in steps of 1$\sigma$ (AK03 and AzTEC1) or 3$\sigma$ (AzTEC5, J1000+0234 and Vd-17871). Different components considered for each source are circled and labeled in the RGB image and potential additional companions are marked with an arrow. The J1000+0234 component confirmed at a lower redshift is labeled with an X. The ALMA beam size is shown at the bottom right corner. North is up, East is to the left, and the images have a size of 5\arcsec$\times$5\arcsec.}
\label{fig:hstalma_multi}
\end{figure*}

The high spatial resolution of the \textit{HST} and ALMA data (0\farcs18 FWHM versus a median synthesized beam size of 0\farcs30$\times$0\farcs27, respectively) allows for detailed studies of the distributions of both obscured and unobscured star formation in the galaxies. The \textit{HST} $F814W$, $F125W$ and $F160W$ images sample the rest-frame stellar UV, which traces un-extincted to moderately extincted star formation, and ALMA band 7 ($\sim 870$\,$\mu$m) samples the rest-frame FIR dust continuum (at $\sim 160$\,$\mu$m for $z = 4.5$), which traces highly obscured star formation. In Figure~\ref{fig:hstalma_multi} we compare these two complementary probes for the objects observed in our \textit{HST} program (all except AzTEC/C159). The \textit{HST} images were PSF-matched and smoothed as described in Sections~\ref{subsec:psfmatch} and \ref{subsec:smooth}.

Qualitatively, the comparison of \textit{HST} and ALMA images suggests important differences in the morphologies. The rest-frame UV stellar emission appears extended and irregular, whereas the rest-frame FIR dust continuum appears very compact. Recently, \citet{2016ApJ...833..103H} found similar results by comparing stellar morphologies from \textit{HST} $F160W$ and ALMA 870\,$\mu$m images in a sample of 16 SMGs at a median redshift of $z \sim 2.5$. \citet{2015ApJ...799..194C} presented similar results regarding the stellar component at rest-frame optical in a larger sample of 48 SMGs at $z = 1$--3.

AK03, AzTEC5, J1000+0234 and Vd-17871 show evidence of two major neighboring components in the rest-frame UV. According to the available spectroscopic redshifts, or photometric redshifts compatible within the uncertainties when we lack spectroscopic confirmation, these components are consistent with being at the same redshift (see Table~\ref{tab:prop}). They also show irregularities and features connecting them (see Figure~\ref{fig:hstalma_multi}). Therefore, it seems very plausible that they are interacting and merging. In addition, AzTEC1 displays a secondary fainter companion towards the North detected in all the three \textit{HST} bands and \textit{Spitzer}/IRAC. Furthermore, AK03, AzTEC5 and J1000+0234 show additional low S/N companions detected also in all the \textit{HST} bands (marked with arrows in Figure~\ref{fig:hstalma_multi}).

All together the full sample is consistent with being multiple component interacting systems. In Section~\ref{sec:mstar} we discuss the stellar mass estimates for the different components of each source. Being able to distinguish the components in the lower resolution datasets, specially in the case of the IRAC bands that trace the rest-frame optical, we obtain stellar masses that are large enough to support the merger scenario as opposed to patches of a single disk or other form of highly extincted single structure.

The compact rest-frame FIR emission, tracing the bulk of the star formation in the system, is always associated with the reddest UV component, but often spatially offset, and not coinciding with the reddest part of the galaxy. This lack of spatial coincidence between the UV and FIR emission is explored further in Section~\ref{sec:dust}.

There are no additional sub-mm detections within the ALMA primary beam at the current sensitivity, and thus, we discard equally bright (close to the phase center) or brighter (away from the phase center) companion DSFGs at distances larger than those showed in the 5\arcsec$\times$5\arcsec images in Figure~\ref{fig:hstalma_multi}.

\subsection{UV Stellar Components} \label{subsec:uv_comp}

In this section we provide a detailed discussion of the individual systems and their subcomponents detected in the \textit{HST} data (see Figure~\ref{fig:hstalma_multi} and Table~\ref{tab:prop}).

\textit{AK03}: This system has two main UV components separated by $\sim 1\arcsec$ (AK03-N and AK03-S), with the $F125W$ image suggesting a bridge connecting the two at an integrated $\rm{S/N} = 2.4$. The spectroscopic confirmation refers to AK03-N, but AK03-S has a comparable photometric redshift \citep[][see Section~\ref{subsec:mstar_cav}]{smolcic15}. All these may be considered evidence for a merger. The dust continuum emission is associated with AK03-S and shows two very compact emission peaks (unresolved at the current resolution), whereas AK03-N remains undetected. Therefore, the bulk of the star formation is associated with AK03-S.

\textit{AzTEC1}: The source shows a compact UV component (AzTEC1-S) and a very faint companion source $\sim 2\arcsec$ towards the North, which is detected at $2 < \rm{S/N} < 3$ in all three \textit{HST} bands (AzTEC1-N). Despite the low S/N of this companion feature, being detected in all three bands the probability of being spurious is $\sim 10^{-5}$. More importantly, it is detected at $\rm{S/N} > 3$ in the HSC $r$, $i$ and $z$ bands, and also in \textit{Spitzer}/IRAC data, confirming that it is a real source. We derived a photometric redshift consistent with lying at the same redshift as AzTEC1-S \citep{2015MNRAS.454.3485Y} within the uncertainties (see Section~\ref{subsec:mstar_cav}). The rest-frame FIR emission is also compact and centered on AzTEC1-S.

\textit{AzTEC5}: For this system, three main UV components are detected in all three \textit{HST} bands (AzTEC5-2, AzTEC5-3 and AzTEC5-4) and a fourth component is detected only in $F814W$ (AzTEC5-1). AzTEC5 is the only source in our sample that lacks spectroscopic confirmation, but photometric redshift estimation indicates a plausible solution for all four components at the same redshift (see Section~\ref{subsec:mstar_cav}). The irregular rest-frame UV morphology of AzTEC5-2 and AzTEC5-4, with emission connecting both in $F160W$, is suggestive of an ongoing merger. The rest-frame FIR has three emission peaks. Two bright peaks associated with AzTEC5-1 and AzTEC5-2 respectively, and a fainter peak in between them, which is not detected in any \textit{HST} bands. Besides, the FIR peaks related with AzTEC5-1 and AzTEC5-2 are aligned with the position of two peaks in the IRAC images, suggesting that the bulk of the stellar mass is associated with these two components which are probably merging.

\textit{AzTEC/C159}: As mentioned in Section~\ref{subsec:hst_data} this source was excluded from the \textit{HST} program and remains undetected in the $F814W$ band image, so we do not have any constraints on its UV morphology. The rest-frame FIR emission is compact and associated with detections in the IRAC bands. 

\textit{J1000+0234}: This system has three main UV components. J1000+0234-N and J1000+0234-S are spectroscopically confirmed at the same redshift \citep[][Karim et al., in prep]{2008ApJ...681L..53C,2008ApJ...689L...5S}. J1000+0234-X is a foreground source at $z_{\rm{spec}} = 1.41$ \citep{2008ApJ...681L..53C}. An additional companion is detected West of J1000+0234-S in all the three \textit{HST} bands, but the HSC images show diffuse features rather than a concentrated source, consistent with \citet{2008ApJ...681L..53C}. The North and South components show a connection between them in all the three \textit{HST} bands, suggesting a merger. The rest-frame FIR emission is compact and associated with J1000+0234-N.

\textit{Vd-17871}: This system has two main UV components $\sim 1\farcs5$ apart (Vd-17871-N and Vd-17871-S), both with elongated morphologies. Both North and South components are spectroscopically confirmed at the same redshift \citep[][Karim et al., in prep]{smolcic15}. The compact rest-frame FIR emission is associated with the North component.

\subsection{SED Fitting} \label{subsec:sed_fit}

Having disentangled different stellar components at rest-frame UV wavelengths, we performed photometry in the lower resolution datasets mentioned in Section~\ref{subsec:add_data}, aiming at fitting the resulting spectral energy distributions (SEDs) to constrain stellar masses for every major stellar component (see Table~\ref{tab:prop}), corresponding to those encircled in Figure~\ref{fig:hstalma_multi}. In the case of \textit{Spitzer}/IRAC, with a significantly lower resolution, the components appear blended, so it is particularly important to know the number of them to properly deblend the fluxes.

\begin{figure*}
\begin{center}
\includegraphics[width=\textwidth]{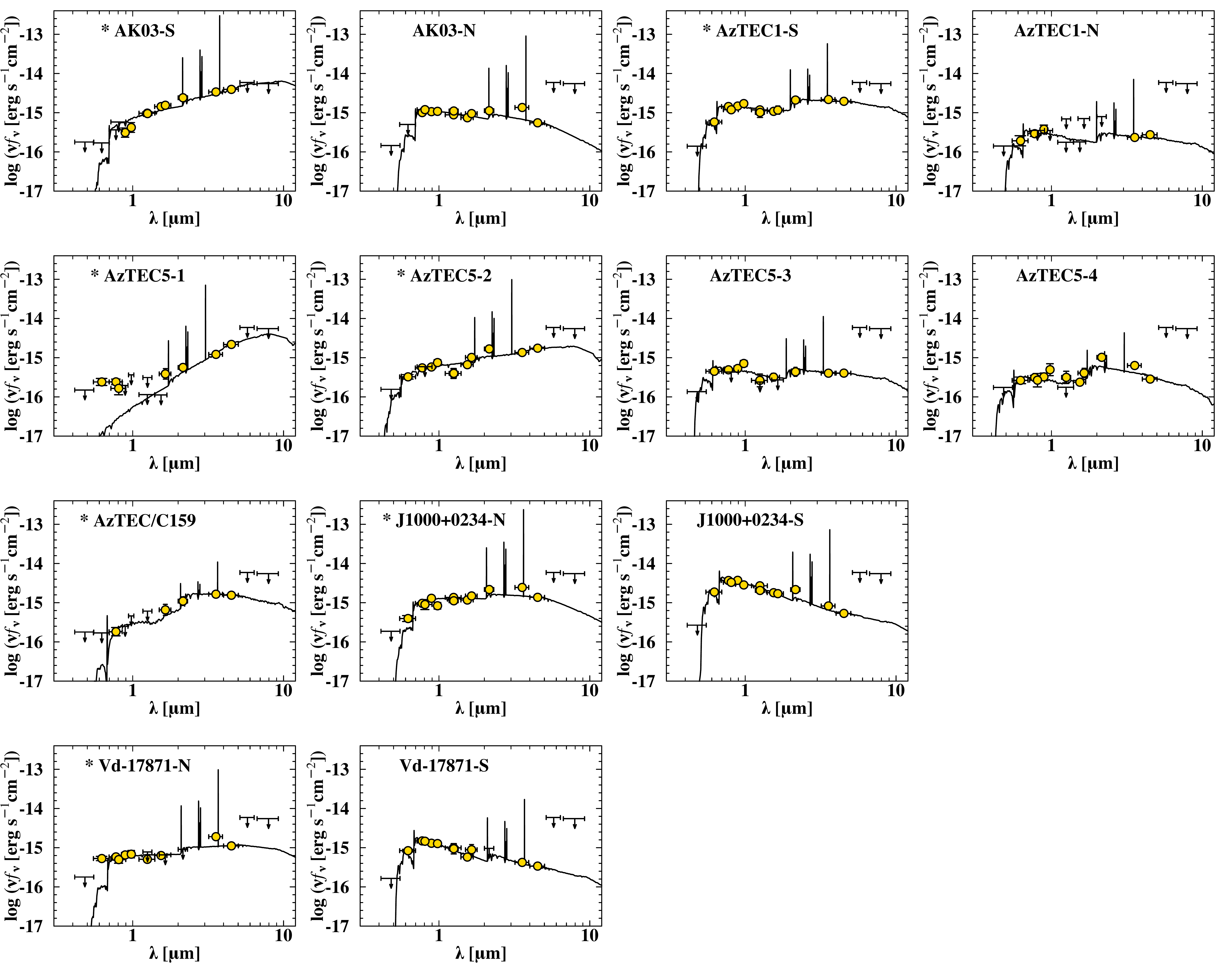}
\caption{SED and best fit for the different stellar components of each object in the sample. Wavelengths are in the observer-frame. Arrows indicate 3$\sigma$ upper limits (5$\sigma$ for the \textit{Spitzer} bands). Component names preceded by a star refer to those with ALMA counterpart.}
\label{fig:seds}
\end{center}
\end{figure*}

From $g$ to $K_{\rm{s}}$ bands the sources are resolved into the stellar components defined from the rest-frame UV \textit{HST} data, appearing unresolved themselves but separated enough, so potential blending is not a concern. To estimate the fluxes in these bands we carried out aperture photometry. The size of the apertures varied for each component and source, being the same across bands, and correspond to those plotted in Figure~\ref{fig:hstalma_multi}. We chose the apertures in the $K_{\rm{s}}$-band to be as large as possible enclosing the component we wanted to study, without overlapping with a neighboring component aperture. We performed aperture corrections for every band. In order to do so, we traced the growth curve of a PSF in the different bands and applied a correction factor to the fluxes accounting for the missing flux outside the aperture. We performed aperture corrections on each band instead of measuring in PSF-matched data to take advantage of the resolution, important for this kind of multiple component systems, that otherwise would be degraded to the lowest-resolution band. The uncertainties in the magnitudes were derived from empty apertures measurements. To assure a good SED fit we only use detections above 3$\sigma$ (upper limits are included in Figure~\ref{fig:seds}).

For the blended \textit{Spitzer}/IRAC 3.6 and 4.5\,$\mu$m images, we employed the magnitudes from a PSF model using the two-dimensional surface brightness distribution fitting algorithm \texttt{GALFIT} \citep{2002AJ....124..266P}. We required at least a 5$\sigma$ detection to perform the fit, which was the case for all source in both 3.6 and 4.5\,$\mu$m bands. The number of PSFs was set to the number of stellar components the source has as defined from the \textit{HST} data and the PSFs centroids were placed at the positions of $K_{\rm{s}}$-band centroids used as priors, allowing a shift in both X and Y axis that turn out to be $< 1$\,pix from the initial positions (IRAC images pixel scale is 0\farcs6\,pix$^{-1}$). The uncertainties in the photometry due to the deblending were calculated by performing a number of realizations varying the centroid coordinates randomly within 1\,pix of the best fit centroid and fixing those coordinates for each realization. Additionally, we checked for detections in the IRAC 5.8 and 8.0\,$\mu$m bands from the S-COSMOS survey \citep{2007ApJS..172...86S}, but the sources are not detected at the required 5$\sigma$ level (upper limits are included in Figure~\ref{fig:seds}).

We fitted the resulting 13-band SEDs ($g$ to 4.5\,$\mu$m, including the three \textit{HST} bands) using LePHARE \citep{1999MNRAS.310..540A,2006A&A...457..841I}. We adopted \citet{2003MNRAS.344.1000B} stellar population synthesis models with emission lines to account for contamination from H$\alpha$ which at the redshift probed in this work is redshifted into the IRAC 3.6\,$\mu$m band. A \citet{2003PASP..115..763C} IMF, exponentially declining star formation histories (SFHs) and a \citet{2000ApJ...533..682C} dust law were assumed. We explored a large parameter grid in terms of SFH e-folding times (0.1\,Gyr–-30\,Gyr), extinction ($0 < A_{V} < 5$), stellar age (1\,Myr--age of the Universe at the source redshift) and metallicity ($Z = 0.004, 0.008$ and 0.02, i.e., solar). The redshift was fixed to the spectroscopic redshift if available or to the photometric redshift if not (see Table~\ref{tab:prop}). In Figure~\ref{fig:seds} we show the derived SEDs, with the fitted models being in good agreement with the data.

\begin{deluxetable*}{lccccccccc}
\tabletypesize{\scriptsize}
\tablecaption{Properties of the Stellar Components. \label{tab:prop}}
\tablehead{\colhead{Source Name} & \colhead{$z_{\rm{spec}}$}\tablenotemark{a} & \colhead{$z_{\rm{phot}}$}\tablenotemark{b} & \colhead{$\beta$} & \colhead{$SFR_{\rm{UV}}$} & \colhead{$L_{\rm{IR}}$}\tablenotemark{c} & \colhead{$SFR_{\rm{IR}}$}\tablenotemark{c} & \colhead{$M_{\rm{dust}}$}\tablenotemark{c} & \colhead{$\log (M_{\rm{*}}/M_{\odot})$}\tablenotemark{d} & \colhead{$M_{\rm{*}}/M_{\rm{*,prim}}$}\tablenotemark{e} \\
\colhead{} & \colhead{} & \colhead{} & \colhead{} & \colhead{[$M_{\odot}$ yr$^{-1}$]} & \colhead{[$10^{12}$ $L_{\odot}$]} & \colhead{[$M_{\odot}$ yr$^{-1}$]} & \colhead{[$10^{8}$ $M_{\odot}$]} & \colhead{} & \colhead{}}
\startdata
$\ast$ AK03-S & \nodata & 4.40 $\pm$ 0.10 & 0.98$_{-0.39}^{+0.49}$ & 25.5 $\pm$ 6.5 & 1.2$_{-0.6}^{+8.1}$ & 120$_{-60}^{+810}$ & 50$_{-35}^{+29}$ & 10.76$_{-0.08}^{+0.08}$ & \nodata \\
AK03-N & 4.747 & 4.75$_{-0.07}^{+0.08}$ & -1.73$_{-0.16}^{+0.16}$ & 53.4 $\pm$ 4.1 & \nodata & \nodata & \nodata & 9.55$_{-0.06}^{+0.06}$ & 1:16.2 \\
$\ast$ AzTEC1-S & 4.3415 & \nodata & -1.16$_{-0.17}^{+0.18}$ & 45.0 $\pm$ 4.0 & 24.0$_{-6.2}^{+8.3}$ & 2400$_{-620}^{+830}$ & 50$_{-10}^{+13}$ & 10.58$_{-0.10}^{+0.10}$ & \nodata \\
AzTEC1-N & \nodata & 3.77$_{-0.22}^{+0.32}$ & -2.6$_{-1.1}^{+1.0}$ & 8.5 $\pm$ 3.3 & \nodata & \nodata & \nodata & 9.56$_{-0.20}^{+0.16}$ & 1:10.5 \\
$\ast$ AzTEC5-1 & \nodata & \nodata & $ < -3.2$\tablenotemark{f} & $ > 4.2$\tablenotemark{f} & 7.9$_{-2.0}^{+1.6}$\tablenotemark{h} & 790$_{-200}^{+160}$\tablenotemark{h} & 9.5$_{-2.5}^{+2.0}$\tablenotemark{h} & 10.40$_{-0.12}^{+0.16}$ & \nodata \\
$\ast$ AzTEC5-2 & \nodata & 3.63$_{-0.15}^{+0.14}$ & 1.6$_{-1.2}^{+1.3}$ & 2.1 $\pm$ 3.7 & 13.2$_{-3.4}^{+2.7}$\tablenotemark{h} & 1320$_{-340}^{+270}$\tablenotemark{h} & 15.8$_{-4.1}^{+3.3}$\tablenotemark{h} & 9.92$_{-0.10}^{+0.10}$ & 1:3.0 \\
AzTEC5-3 & \nodata & 4.02$_{-0.08}^{+0.08}$ & $-$0.25$_{-0.78}^{+1.1}$ & 3.8 $\pm$ 2.6 & \nodata & \nodata & \nodata & 9.78$_{-0.10}^{+0.08}$ & 1:4.2 \\
AzTEC5-4 & \nodata & 3.66$_{-0.43}^{+0.40}$ & $-$1.12$_{-0.52}^{+0.66}$ & 5.2 $\pm$ 2.1 & \nodata & \nodata & \nodata & 9.59$_{-0.06}^{+0.08}$ & 1:6.5 \\
$\ast$ AzTEC/C159 & 4.567 & \nodata & $ > -1.2$\tablenotemark{g} & $ < 33$\tablenotemark{g} & 7.4$_{-1.7}^{+2.1}$ & 740$_{-170}^{+210}$ & 25.0$_{-5.0}^{+6.0}$ & 10.65$_{-0.08}^{+0.08}$ & \nodata \\
$\ast$ J1000+0234-N & 4.539 & \nodata & $-$1.01$_{-0.32}^{+0.39}$ & 52.6 $\pm$ 8.5 & 4.4$_{-3.2}^{+12}$ & 440$_{-320}^{+1200}$ & 50$_{-34}^{+110}$ & 10.14$_{-0.08}^{+0.08}$ & \nodata \\
J1000+0234-S & 4.547 & 4.48$_{-0.03}^{+0.03}$ & $-$2.04$_{-0.11}^{+0.12}$ & 147.6 $\pm$ 7.4 & \nodata & \nodata & \nodata & 9.16$_{-0.08}^{+0.06}$ & 1:9.5 \\
$\ast$ Vd-17871-N & 4.621 & 4.49$_{-0.03}^{+0.04}$ & $-$0.59$_{-0.31}^{+0.35}$ & 22.1 $\pm$ 4.0 & 11.2$_{-2.3}^{+2.9}$ & 1120$_{-230}^{+290}$ & 12.6$_{-2.6}^{+3.2}$ & 10.04$_{-0.10}^{+0.10}$ & \nodata \\
Vd-17871-S & 4.631 & 4.41$_{-0.09}^{+0.08}$ & $-$2.27$_{-0.23}^{+0.22}$ & 59.3 $\pm$ 5.5 & \nodata & \nodata & \nodata & 9.49$_{-0.30}^{+0.18}$ & 1:3.5 \\
\enddata
\tablenotetext{}{Component names preceded by $\ast$ refer to those with ALMA counterpart.}
\tablenotetext{a}{Spectroscopic redshift references: AK03-N from Ly$\alpha$ by \citet{smolcic15}; AzTEC1-S from [\ion{C}{2}], also $^{12}$CO(4-3) and $^{12}$CO(5-4), by \citet{2015MNRAS.454.3485Y}; AzTEC/C159 from [\ion{C}{2}] by Karim et al. (in prep), see also $^{12}$CO(2-1) and $^{12}$CO(5-4) by \citet{2017arXiv171010181J}, and Ly$\alpha$ by \citet{smolcic15}; J1000+0234-N from [\ion{C}{2}] by Karim et al. (in prep), see also $^{12}$CO(4-3) by \citet{2008ApJ...689L...5S}, and Ly$\alpha$ by \citet{2008ApJ...681L..53C}; J1000+0234-S from Ly$\alpha$ by \citet{2008ApJ...681L..53C}; Vd-17871-N from [\ion{C}{2}] by Karim et al. (in prep), see also \citet{smolcic15}; Vd-17871-S from Ly$\alpha$ by Karim et al. (in prep), see also \citet{smolcic15}.}
\tablenotetext{b}{Photometric redshift references: AK03-S from \citet{smolcic15} who found a $z_{\rm{phot}} = 4.40 \pm 0.10$ or $z_{\rm{phot}} = 4.65 \pm 0.10$, depending on the template used; AzTEC1-N calculated in this work, where the uncertainties correspond to the 1$\sigma$ percentiles of the maximum likelihood distribution and the redshift distribution spans over the range $3.2 < z < 4.7$; AK03-N, AzTEC5-2, AzTEC5-3 and AzTEC5-4 from the 3D-\textit{HST} survey catalog \citep{2016ApJS..225...27M,2012ApJS..200...13B,2014ApJS..214...24S}; J1000+0234-S, Vd-17871-N and Vd-17871-S from the COSMOS2015 catalog \citep{2016ApJS..224...24L}; AzTEC/C159, J1000+0234-N and AzTEC5-1 have no counterpart in the COSMOS2015 catalog. For both 3D-\textit{HST} and COSMOS2015 estimates the listed uncertainties correspond to the 1$\sigma$ percentiles.}
\tablenotetext{c}{From \citet{smolcic15} (\citet{2014ApJ...782...68T} for AzTEC5) FIR SEDs covering 100\,$\mu$m--1.1\,mm updated with new 850\,$\mu$m fluxes from the S2COSMOS/SCUBA2 survey (Simpson et al., in prep). $L_{\rm{IR}}$ (integrated from rest-frame 8--1000\,$\mu$m) and $M_{\rm{dust}}$ are infered using the \citet{2007ApJ...657..810D} dust model, then $SFR_{\rm{IR}}$ is calculated using the $L_{\rm{IR}}$ to $SFR_{\rm{IR}}$ conversion from \citet{kennicutt98} for a Chabrier IMF.}
\tablenotetext{d}{Stellar mass uncertainties do not reflect systematics due to the SED fitting assumptions (i.e., stellar population synthesis models, IMF or SFH).}
\tablenotetext{e}{Stellar mass ratio between the quoted and the most massive components ($M_{\rm{*,prim}}$).}
\tablenotetext{f}{Limits from detection in $F814W$ and upper limits in $F125W$ and $F160W$.}
\tablenotetext{g}{Limits from UltraVISTA DR3 photometry.}
\tablenotetext{h}{AzTEC5-1 accounts for 30\% and AzTEC5-2 for 50\% of the total values for this source following our GALFIT ALMA continuum images modeling (see Section~\ref{subsec:fir_size}).}
\end{deluxetable*}

\section{Dust Absorption and Emission} \label{sec:dust}

\subsection{Spatially Resolved UV Slopes} \label{subsec:betamap}

At the redshift of the galaxies, the three \textit{HST} bands trace the rest-frame UV continuum. This makes it possible to directly determine their spatially resolved UV slopes ($\beta$). 

\begin{figure*}
\begin{center}
\includegraphics[width=\textwidth]{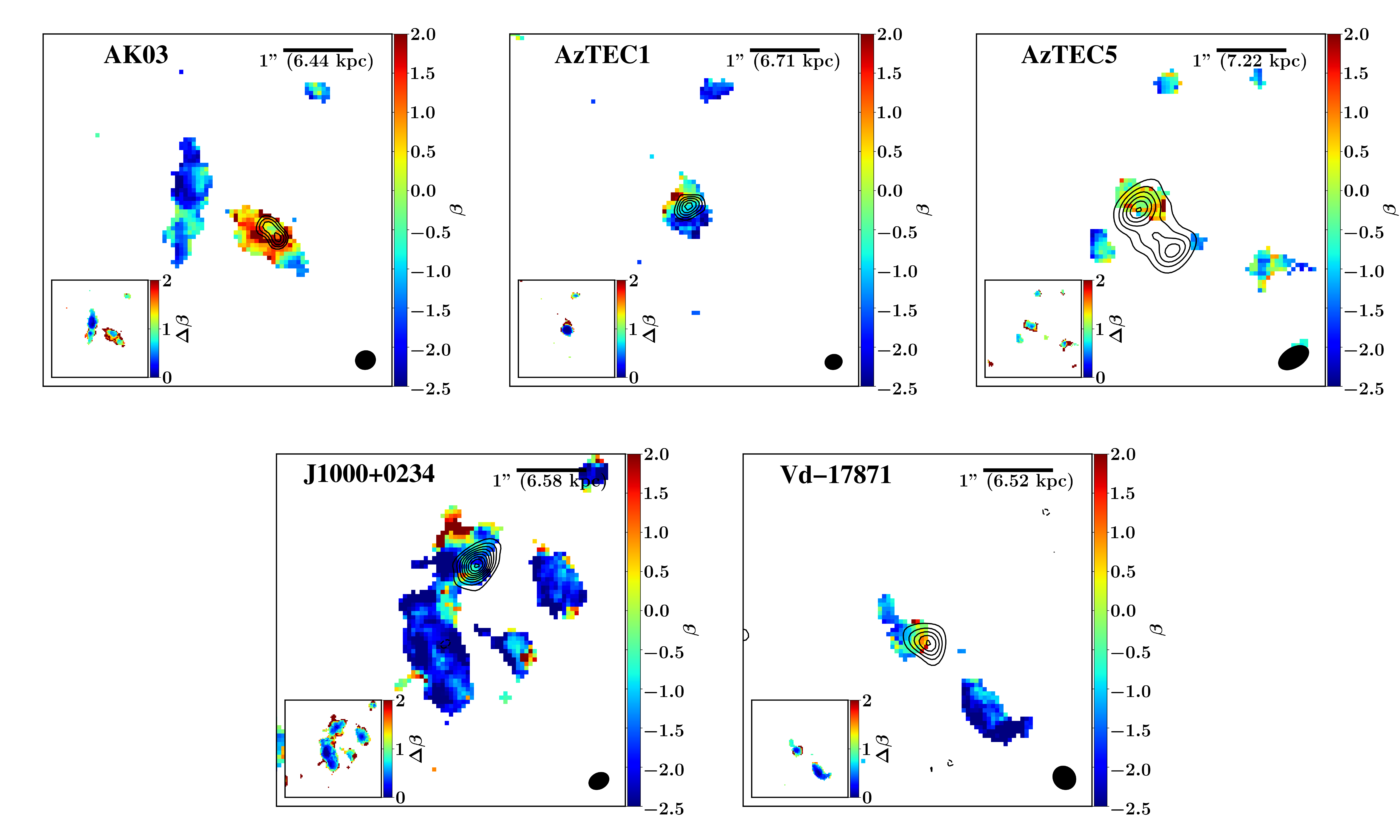}
\caption{UV continuum slope maps of the five sources observed with \textit{HST}. ALMA band 7 ($\sim 870$\,$\mu$m) contours are overlayed starting at $\pm3\sigma$ in steps of 1$\sigma$ (AK03 and AzTEC1) or 3$\sigma$ (AzTEC5, J1000+0234 and Vd-17871). The error map is shown in the bottom left corner and the ALMA beam size at the bottom right corner of each panel. North is up, East is to the left, and the images have a size of 5\arcsec$\times$5\arcsec.}
\label{fig:betamap}
\end{center}
\end{figure*}

In Figure~\ref{fig:betamap} we present $\beta$ maps, constructed by fitting a linear slope to pixels which have $\rm{S/N} > 2$ detections in at least two smoothed images (see Section~\ref{subsec:smooth}). The 1$\sigma$ uncertainty maps (inserts) were constructed by computing $\beta$-values in $\sim 10000$ realizations of the data, varying in each realization the measured pixel flux values within their uncertainties. Note that the pixel size is 0\farcs06, but the PSF FWHM is 0\farcs18. Consequently, spatially independent regions are those separated by at least 3 pixels. Since the UV slope maps were obtained using at least two detections in the \textit{HST} bands, we see more clearly the presence of faint companions towards the North in AK03, AzTEC1, AzTEC5 and J1000+0234, as mentioned in Section~\ref{sec:morph}.

In general, the objects present blue UV slopes, but the values are not homogeneous over the extent of the galaxies. The color gradients could be caused by structure in the distribution of dust, stellar age or metallicity. The relative importance of these cannot be disentangle with the available data but we expect a patchy dust distribution to be the dominant cause. However, as most of the extent of the rest-frame UV emission is not detected in our ALMA observations, revealing the underlying dust structure in emission would require deeper observations.

The rest-frame FIR dust emission is in all cases associated with the reddest components. These components show evidence of gradients in their UV slopes. AK03-S is redder towards the North-East and bluer at the South-West. J1000+0234-N has an extended redder feature at the North-East. Vd-17871-N is slightly redder towards the South-West direction and bluer towards the North-East. In AzTEC1-N the red-to-blue gradient goes along the North-South axis. These color gradients may be due to a star formation gradient with higher dust content towards the redder areas. Another possibility could be close mergers between red and blue galaxies.

In AK03-S, two close FIR peaks are detected. At the current resolution and sensitivity and without dynamical information, we cannot determine whether these are part of a larger dynamical structure like a clumpy disk or remnants of a past interaction/merger. Note that in AzTEC5 we were unable to constrain the resolved UV slope of AzTEC5-1 since it is only detected in $F814W$, suggesting a extremely high extinction with strong rest-frame FIR emission, but also a very blue rest-frame UV component.

The bluer components in all five systems remain undetected in the ALMA continuum. This indicates less dusty star formation.

Spatially integrated values for the UV slopes (see Table~\ref{tab:prop}) show a median and median absolute deviation of $\beta = -0.59 \pm 0.57$ for the components associated with the ALMA continuum emission (namely AK03-S, AzTEC1-S, AzTEC5-2, J1000+0234-N, Vd-17871-N and excluding the AzTEC5-1 and AzTEC/C159 upper limits). The rest of the components are bluer, with $\beta = -1.73 \pm 0.54$ consistent with estimates of Lyman break galaxies (LBGs) at similar redshift \citep[e.g.,][]{2009ApJ...705..936B,2012A&A...540A..39C,2012ApJ...756..164F}. By performing the photometry over larger apertures, enclosing all the components per source, we derive $\beta = -0.91 \pm 0.85$, which is in between the derived values for the red and blue components.

Having identified which UV components are associated with the dust continuum emission, we can relate the star formation rate (SFR) in the infrared ($SFR_{\rm{IR}}$), tracing the obscured star formation, with that in the ultraviolet ($SFR_{\rm{UV}}$), probing the unobscured star formation. The former was obtained from the FIR SEDs presented in \citet{smolcic15} (\citet{2014ApJ...782...68T} for AzTEC5) covering 100\,$\mu$m--1.1\,mm updated with new 850\,$\mu$m fluxes from the S2COSMOS/SCUBA2 survey (Simpson et al., in prep). The procedure is the following: The FIR SED is modeled using the \citet{2007ApJ...657..810D} dust model (DL07) \citep[e.g.,][]{2012ApJ...760....6M,2017A&A...603A..93M,2016A&A...587A..73B}; $L_{\rm{IR}}$ is calculated by integrating the best fit to the SED in the range 8--1000\,$\mu$m; and then $SFR_{\rm{IR}}$ is obtained using the $L_{\rm{IR}}$ to $SFR_{\rm{IR}}$ conversion from \citet{kennicutt98} for a Chabrier IMF. $SFR_{\rm{UV}}$ was calculated employing \citet{2007ApJS..173..267S} prescription, that relates $L_{\rm{UV}}$ to $SFR_{\rm{UV}}$, for a Chabrier IMF. Note that $SFR_{\rm{UV}}$ derived this way corresponds to the observed value, i.e., not corrected from extinction. The total SFR can be accounted by adding both infrared and ultraviolet estimates ($SFR_{\rm{IR+UV}}$). Not suprisingly the star formation is dominated by $SFR_{\rm{IR}}$, with $SFR_{\rm{UV}}$ only contributing at the level of 2--20\% to the total SFR ($SFR_{\rm{IR+UV}}$), in agreement with other previous works comparing obscured and unobscured star formation in starburst galaxies \citep[e.g.,][]{2017ApJ...838L..18P} and galaxies with similar stellar mass \citep[e.g.,][]{2017ApJ...850..208W}.

In relation with the $L_{\rm{IR}}$ and $SFR_{\rm{IR}}$ estimates it is important to consider whether an important fraction of the infrared emission could be related with active galatic nuclei (AGN) activity. As reported in \citet{smolcic15}, none of the sources is detected in the X-ray catalog in the COSMOS field \citep[\textit{Chandra} COSMOS Legacy Survey;][]{2016ApJ...819...62C,2016ApJ...817...34M}. In terms of the radio emission \citet{smolcic15} studied the infrared-radio correlation of the sample, which show a discrepancy when compared with low-redshift star-forming galaxies due to a mild radio excess. This excess would be in line with studies showing an evolving infrared-radio ratio depending on the age on the starburst. In any case, while many SMGs host AGN, their $L_{\rm{IR}}$ is dominated by the star formation with the AGN contribution being $< 33\%$ \citep[e.g.,][]{2008ApJ...675.1171P,2014ApJ...786...31R}. This translates into a maximum overestimation in the $SFR_{\rm{IR}}$ of 33\%, below the $SFR_{\rm{IR+UV}}$ sample scatter.

\subsection{FIR Sizes} \label{subsec:fir_size}

We measured the sizes of the rest-frame FIR dust continuum emission by modeling the ALMA continuum images using \texttt{GALFIT}. S\'ersic and PSF profiles were fitted to compare both resolved and unresolved modeling of the objects. The only object that was better fitted by a point source than a S\'ersic model (and thus unresolved) is AK03. For this galaxy we derived an upper limit on the size from the PSF.

For the rest of the galaxies we fitted models with the S\'ersic index fixed to $n = 0.5$, 1 and 4, corresponding to a gaussian, exponential disk and de Vaucouleurs profiles, respectively, and also leaving the index free. The size of the emitting regions was obtained through the effective radius of the models ($r_{\rm{e}}$). We cannot constrain which S\'ersic index better explains the data at the current resolution and S/N. From higher resolution observations \citet{2016ApJ...833..103H} found a median S\'ersic index of $n = 0.9 \pm 0.2$ for a sample of 15 SMGs and concluded that the dust emission follows an exponential disk profile. Motivated by this, we fixed $n = 1$ to report the rest-frame FIR sizes for our sample in Table~\ref{tab:fir_size}. We also performed fits varying the axis ratio ($b/a$) and found that no particular value with $b/a \geq 0.3$ fitted the data better than others, so we fixed it to the circular value $b/a = 1$. We take into account the possible systematic errors associated with the assumed S\'ersic index and axis ratio in the listed effective radii errors. These were computed by adding in quadrature the statistical uncertainty from \texttt{GALFIT} for the circular disk model and the difference between this model and the full range of models with varying $n$ and $b/a$. Therefore, the uncertainties conservatively account for the inability of the data to robustly constrain the detailed shape of the surface brightness profiles. We note that the ALMA continuum fluxes are consistent with the 850\,$\mu$m fluxes from the S2COSMOS/SCUBA2 survey (Simpson et al., in prep), thus there is no evidence for resolved-out or missing flux that could affect the size estimates.

Finally, we cross-checked the results analyzing the data directly in the ($u,v$) plane employing \texttt{UVMULTIFIT} \citep{2014A&A...563A.136M} following the procedure described in \citet{2017ApJ...850...83F}. In this case for a direct comparison with the \texttt{GALFIT} image plane fits we also assumed a circular disk model (to obtain secure results, we omit AK03 and AzTEC5 for this comparison as they show two and three components respectively in our ALMA continuum images). We find that these estimates are in agreement with the results derived in the reconstructed images using \texttt{GALFIT} (see Table~\ref{tab:fir_size}). In the following we use the estimates derived from \texttt{GALFIT} for further calculations.

\begin{deluxetable*}{lcccccc}
\tabletypesize{\scriptsize}
\tablecaption{Rest-frame FIR Sizes. \label{tab:fir_size}}
\tablehead{\colhead{Source Name}\tablenotemark{a} & \colhead{$r_{\rm{e}}^{\rm{GALFIT}}$} & \colhead{$r_{\rm{e}}^{\rm{GALFIT}}$} & \colhead{$r_{\rm{e}}^{\rm{UVMULTIFIT}}$} & \colhead{$r_{\rm{e}}^{\rm{UVMULTIFIT}}$} & \colhead{$\Sigma_{\rm{SFR}}$\tablenotemark{b}} & \colhead{$\Sigma_{\rm{dust}}$\tablenotemark{b}} \\
\colhead{} & \colhead{[pc]} & \colhead{[$''$]} & \colhead{[pc]} & \colhead{[$''$]} & \colhead{[$M_{\odot}$ yr$^{-1}$ kpc$^{-2}$]} & \colhead{[10$^9$ $M_{\odot}$ kpc$^{-2}$]}}
\startdata
AK03-S\tablenotemark{c}       & $ < 520$ & $ < 0.08$ & \nodata & \nodata & $ > 3.4$ & $ > 1.4$ \\
\nodata             & $ < 520$ & $ < 0.08$ & \nodata & \nodata & $ > 3.8$ & $ > 1.6$ \\
AzTEC1-S     & 900$_{-290}^{+480}$ & 0.13$_{-0.04}^{+0.07}$ & 940 $\pm$ 70 & 0.14 $\pm$ 0.01 & 480$_{-340}^{+540}$ & 1.0$_{-0.7}^{+1.1}$ \\
AzTEC5-1\tablenotemark{d}     & 300$_{-130}^{+90}$ & 0.04$_{-0.02}^{+0.01}$ & \nodata & \nodata & 1260$_{-1200}^{+870}$ & 1.7$_{-1.5}^{+1.1}$ \\
\nodata             & 560$_{-360}^{+120}$ & 0.08$_{-0.05}^{+0.02}$ & \nodata & \nodata & 250$_{-140}^{+330}$ & 0.33$_{-0.43}^{+0.16}$ \\
AzTEC5-2     & 700$_{-390}^{+180}$ & 0.10$_{-0.05}^{+0.03}$ & \nodata & \nodata & 390$_{-440}^{+240}$ & 0.51$_{-0.58}^{+0.30}$ \\
AzTEC/C159   & 460$_{-240}^{+60}$ & 0.07$_{-0.04}^{+0.01}$ & 590 $\pm$ 70 & 0.09 $\pm$ 0.01 & 570$_{-610}^{+220}$ & 1.9$_{-0.7}^{+2.0}$ \\
J1000+0234-N & 700$_{-100}^{+120}$ & 0.11$_{-0.02}^{+0.02}$ & 660 $\pm$ 70 & 0.10 $\pm$ 0.01 & 150$_{-110}^{+380}$ & 1.6$_{-1.2}^{+3.5}$ \\
Vd-17871-N   & 370$_{-210}^{+80}$ & 0.06$_{-0.03}^{+0.01}$ & 650 $\pm$ 70 & 0.10 $\pm$ 0.01 & 1300$_{-1500}^{+670}$ & 5.8$_{-7.7}^{+4.2}$ \\
\enddata
\tablenotetext{a}{Names refer to the stellar component associated with the FIR emission.}
\tablenotetext{b}{Defined as $\Sigma_{\rm{SFR}} = 0.5SFR/\pi ({r_{\rm{e,circ}}^{\rm{GALFIT}}})^2$ and $\Sigma_{\rm{dust}} = 0.5M_{\rm{dust}}/\pi ({r_{\rm{e,circ}}^{\rm{GALFIT}}})^2$.}
\tablenotetext{c}{Limits from the PSF referring to each one of the two emitting regions.}
\tablenotetext{d}{The three values of AzTEC5 allude to the three resolved emitting regions from West to East.}
\end{deluxetable*}

The median and median absolute deviation of the size estimate for our sample are then $r_{\rm{e}} = 0.70 \pm 0.29$\,kpc at $\sim 870$\,$\mu$m, which corresponds $\sim 160$\,$\mu$m rest-frame at $z = 4.5$ (excluding AK03 upper limits and only considering the brightest peak in AzTEC5, associated with AzTEC5-2). This result is in good agreement with \citet{2015ApJ...810..133I}, who found similar compact sizes of $r_{\rm{e}} = 0.67_{-0.14}^{+0.13}$\,kpc for a sample of 13 1.1\,mm-selected SMGs at a comparable redshift $3 < z < 6$. \citet{2017arXiv170904191O} presented an average value of $r_{\rm{e}} = 0.91 \pm 0.26$\,kpc (converting the reported FWHM into a circularized effective radius) in a sample of 44 DSFGs at $z \sim 4$--6 observed at $\sim 870$\,$\mu$m and selected as \textit{Herschel} 500\,$\mu$m risers (SED rise from 250\,$\mu$m to 500\,$\mu$m). On the other hand, the typical sizes derived for SMGs at a median redshift of $z \sim 2.5$ were reported to be $r_{\rm{e}} = 1.8 \pm 0.2$\,kpc from \citet{2016ApJ...833..103H} and also $r_{\rm{e}} = 1.2 \pm 0.1$\,kpc from \citet{2015ApJ...799...81S}, both targetting $\sim 870$\,$\mu$m. This suggest that SMGs may be more compact at $z > 3$ than at $z < 3$ \citep[e.g.,][]{2017ApJ...850...83F,2017arXiv170904191O}. Other individual sources at $z > 4$ also point towards very compact dust continuum emission \citep[e.g.,][]{2013Natur.496..329R,2014ApJ...796...84R,2016ApJ...816L...6D} and also pairs of compact interacting starburst galaxies detected in gas and dust continuum which suggests a gas-rich major merger \citep[e.g.,][]{2016ApJ...827...34O,2017ApJ...850....1R}. \citet{2016ApJ...826..112S} found no evidence for a difference in the size distribution of lensed DSFGs compared to unlensed samples from a sample of 47 DSFGs at $z = 1.9$--5.7. Our results are also similar to the compact morphologies of local ULIRGs \citep[$r_{\rm{e}} = 0.5$\,kpc,][]{2016A&A...591A.136L} at 70\,$\mu$m rest-frame. We note that caution should be exercised when comparing samples tracing different rest-frame FIR wavelenghts and based on different selection methods. Another caveat for a fair comparison is the stellar mass, since more massive galaxies are typically larger \citep[e.g.,][]{2014ApJ...788...28V}.

From the $SFR_{\rm{IR}}$ obtained for these sources (see Table~\ref{tab:prop}) and their rest-frame FIR sizes, we calculated the SFR surface density ($\Sigma_{\rm{SFR}} = 0.5SFR/\pi r_{\rm{e,circ}}^2$, see Table~\ref{tab:fir_size}). Ranging from $\Sigma_{\rm{SFR}} = 150$--1300\,$M_{\odot}$ yr$^{-1}$ kpc$^{-2}$ (excluding AK03 lower limits), the most extreme cases are AzTEC1, AzTEC5-1 and Vd-17871-N, but the last two are poorly constrained due to the large uncertainty on their sizes. At such extreme values, they are candidates for Eddington-limited starbursts \citep[$\Sigma_{\rm{SFR}} \sim 1000$\,$M_{\odot}$ yr$^{-1}$ kpc$^{-2}$,][]{2011ApJ...727...97A,2015ApJ...799...81S}.

\subsection{UV/FIR Spatial Disconnection} \label{subsec:uvfir_discon}

The dust masses derived for this sample are very high at $\sim 10^9$\,$M_{\odot}$ (see Table~\ref{tab:prop}). Dust masses are a free parameter in the DL07 model employed, controlling the normalization of the SED. In terms of the dust opacity, DL07 assumes optically thin dust ($\tau << 1$) at all wavelengths \citep[e.g.,][]{2012ApJ...760....6M,2017A&A...603A..93M,2016A&A...587A..73B}. Very high dust masses combined with the small sizes derived for the dust emitting regions implies very high dust mass surface densities ($\Sigma_{\rm{dust}} = 0.5M_{\rm{dust}}/\pi r_{\rm{e,circ}}^2$), with values ranging $\Sigma_{\rm{dust}} = 0.33$--$5.8 \times 10^{9}$\,$M_{\odot}$ kpc$^{-2}$ (see Table~\ref{tab:fir_size}), and consequently, very high extinction.

We calculated the expected extinction assuming that the dust is distributed in a sheet with uniform density. We inferred the mean extinction from the dust mass surface density-to-extinction ratio ($\Sigma_{\rm{dust}}/A_{V}$). To calculate $\Sigma_{\rm{dust}}/A_{V}$ we assumed a gas-to-dust mass ratio (GDR) appropriate for SMGs of $\rm{GDR} = 90$ \citep{2014MNRAS.438.1267S}, and the gas surface number density-to-extinction ratio $N_{\rm{H}}/A_{V} = 2.2 \times 10^{21}$\,cm$^{-2}$ mag$^{-1}$ \citep{2011A&A...533A..16W}. Therefore, $\Sigma_{\rm{dust}}/A_{V} = (N_{\rm{H}}/A_{V}) \cdot m_{\rm{H}} / \rm{GDR} = 2.44$\,$M_{\odot}$ pc$^{-2}$ mag$^{-1}$. With this number the mean extinction is $\langle A_{V} \rangle = \Sigma_{\rm{dust}}/2.44$. The values for our sample are extreme $\langle A_{V} \rangle = 130$--2400\,mag, even when the numbers are halved to account for the dust behind the sources \citep[see also][]{2017ApJ...844L..10S}.

Comparing Figures~\ref{fig:hstalma_multi} and \ref{fig:betamap} we see that while the dust emission is always associated with the reddest (likely most dust-extincted) component, in most cases it is not centered on the reddest part of that component (with the possible exception of Vd-17871). This suggests that the extinction seen in emission and absorption are disconnected, consistent with the expected extreme $A_{V}$ which implies that no emission can escape at any wavelength.

The fact that we do see blue UV emission at the peak of the dust emission suggests that a fraction of the light is able to escape due to a clumpy dust distribution and/or that the dust and stars are seen in different projections, e.g., the stars responsible for the UV emission could be in front of the dusty starbursts.

In any case it is clear that the rest-frame UV and FIR emissions are spatially disconnected and originate from a different physical region. This implies that the dust as seen in absorption from the UV slope inhomogeneities in Figure~\ref{fig:betamap} is not tracing the dust seen in emission from the ALMA continuum.

\subsection{IRX-$\beta$ Plane} \label{subsec:irxbeta}

The infrared-to-ultraviolet luminosity ratio, commonly refered as infrared excess ($\rm{IRX} = L_{\rm{IR}}/L_{\rm{UV}}$), is known to correlate with the UV continuum slope ($\beta$). This so-called Meurer relation \citep[][M99 relation hereafter]{1999ApJ...521...64M} is well established for normal star-forming galaxies \citep[e.g.,][]{2011ApJ...726L...7O,2012ApJ...755..144T,2014ApJ...796...95C}. Its origin is thought to be that galaxies get redder as the dust absorbs the rest-frame UV emission and re-radiates it at infrared wavelengths. For galaxies on the relation, the amount of dust absorption can thus be directly inferred from the UV slope. Therefore, in the absence of FIR data, the relation can be used to obtain total extinction-corrected SFR from UV data \citep[e.g.,][]{2009ApJ...705..936B}. Furthermore, this relation physically motivates energy balance codes which require that dust extinction inferred from rest-frame UV--optical SED fits must match the observed emission measured at infrared wavelengths \citep[e.g.,][]{2008MNRAS.388.1595D}.

Spatially unresolved observations have shown that DSFGs do not follow the M99 relation \citep[e.g.,][]{2005ApJ...619L..51B,2010ApJ...715..572H,2014ApJ...796...95C}. Excess of dust and UV/FIR decoupling have been suggested as a possible origin of the offsets by \citet{2010ApJ...715..572H} who showed that the deviation from the nominal M99 relation ($\Delta$IRX) increases with $L_{\rm{IR}}$, but does not correlate with $L_{\rm{UV}}$. Following this argument the authors postulated that a concentration parameter might correlate with $\Delta$IRX as an indicator of the decoupled UV/FIR. \citet{2014ApJ...796...95C} reinforced these results showing also that the deviation from the M99 relation increases with $L_{\rm{IR}}$ above a threshold of $\log(L_{\rm{IR}}/L_{\odot}) > 11.0$. \citet{2017ApJ...847...21F} proposed that the blue colors of sources with high IRX values could be due to holes in the dust cover, tidally stripped young stars or faint blue satellite galaxies. In addition, simulations propose recent star formation in the outskirts and low optical depths in UV-bright regions as plausible explanations of the offset \citep{2017ApJ...840...15S,2018MNRAS.474.1718N}. Simple models placing a dust screen in front of a starburst have been studied to provide a detailed explanation of all the possible effects that might lead to a deviation in the IRX-$\beta$ plane \citep{2017MNRAS.472.2315P}.

\begin{figure}
\begin{center}
\includegraphics[width=\columnwidth]{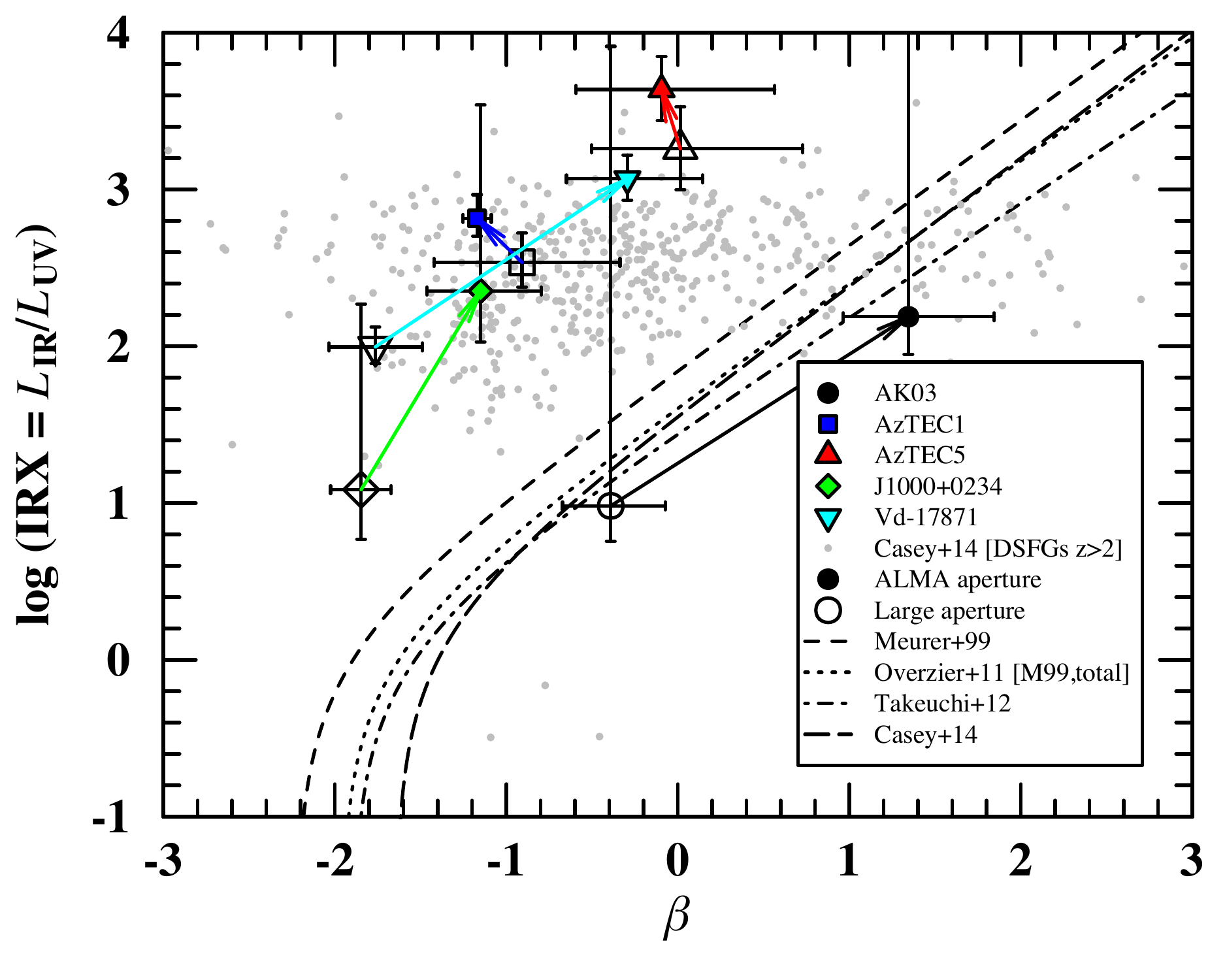}
\caption{IRX-$\beta$ plane. Small filled symbols indicate \textit{HST} photometry performed over the region above the 3$\sigma$ contour in the ALMA image. Large open symbols refer to \textit{HST} photometry performed over a large aperture enclosing all the components of each source. Both small and large symbols are linked with an arrow of the color associated to each object. Small gray dots indicate the \citet{2014ApJ...796...95C} sample of DSFGs at $z > 2$. IRX-$\beta$ relations from the literature include: \citet{1999ApJ...521...64M} (dashed line), corrected M99 relation referred as ``M99, total'' in \citet{2011ApJ...726L...7O} (dotted line), \citet{2012ApJ...755..144T} (dashed-dotted line), \citet{2014ApJ...796...95C} (long-dashed line). Our sample does not follow the M99 relation which, together with their rest-frame UV and FIR morphologies, suggests that the UV and FIR emissions of DSFGs are spatially disconnected.}
\label{fig:irxbeta}
\end{center}
\end{figure}

The sample studied here have infrared luminosities ranging $\log(L_{\rm{IR}}/L_{\odot}) = 12.1$--13.4, above the mentioned threshold $\log(L_{\rm{IR}}/L_{\odot}) > 11.0$, and the spatially resolved rest-frame UV/FIR data make it possible to study the origin of the DSFGs offsets in the IRX-$\beta$ plane (see Figure~\ref{fig:irxbeta}).

To confirm that the galaxies in this sample are representative of previous DSFGs studies in spatially unresolved data, we first derived ultraviolet and infrared luminosities in large apertures enclosing all the components of each source. In Figure~\ref{fig:irxbeta} these measurements are plotted as large open symbols, confirming that the sample does not follow the M99 relation and it is located in the same region as previous spatially unresolved measurements for DSFGs \citep[e.g.,][at $z > 2$]{2014PhR...541...45C}. Second, we take advantage of the spatial resolution to pinpoint the origin of the FIR emission and recalculate the UV luminosity in smaller apertures defined by the 3$\sigma$ contour in the ALMA images (ALMA apertures). In this case both \textit{HST} and ALMA images were PSF-matched as described in Section~\ref{subsec:psfmatch}.

In Figure~\ref{fig:irxbeta} we plot the sample of DSFGs at $z > 2$ from \citet{2014ApJ...796...95C} for comparison. Note that this study employed similar methods to obtain $L_{\rm{IR}}$, $L_{\rm{UV}}$ and the UV slopes as we did: $L_{\rm{IR}}$ by integrating over the wavelength range 8--1000\,$\mu$m and using a single temperature modified greybody plus mid-IR power law, which properly accounts for the warm dust contribution as the DL07 dust model; $L_{\rm{UV}}$ by interpolating the observed photometry to rest-frame 1600\,\AA; and the UV slopes by fitting a power law to the photometry, which is equivalent to our linear fit in magnitude space. Additionally, we include other IRX-$\beta$ relations from the literature: the original M99 and follow-up corrections \citep[e.g.,][]{2011ApJ...726L...7O,2012ApJ...755..144T}, although the methodology they followed to obtain the quantities shown in the IRX-$\beta$ diagram slightly differ from \citet{2014ApJ...796...95C} and ours.

All the galaxies have higher IRX in the ALMA aperture than in the large aperture. This is expected from their smaller extent in the rest-frame FIR compared to that in the rest-frame UV, which effectively lowers the $L_{\rm{UV}}$ contribution to the $L_{\rm{IR}}/L_{\rm{UV}}$ ratio. Furthermore, three of the galaxies have redder UV slopes and two have similar UV slopes in the ALMA apertures compared with the large apertures. Again this can be understood as a result of removing the contribution from the extended irregular UV features and companion satellite galaxies that appear bluer than the dust emitting region detected in ALMA. These results agree with the model proposed in \citet{2017ApJ...847...21F} to explain blue colors of DSFGs with high IRX.

On the other hand, even after accounting for the correction that implies going from the large to the ALMA aperture, our sample does not follow the M99 relation and lies 1.75\,dex (median) above it. However, while the rest-frame FIR dust continuum emission is associated with the reddest component in the mergers, it is in general not centered on the reddest part of the component, and the component is too blue to be consistent with a physical connection between the dust seen in emission and absorption, suggesting that the UV and FIR emissions of DSFGs are spatially disconnected.

This provides morphological and geometrical evidence for the origin of the DSFGs offsets from the M99 relation \citep[see also][]{2017ApJ...846..108C}, being consistent with the extreme extinction expected from the compact and intense dust emission for this sample (see Section~\ref{subsec:uvfir_discon}), implying that UV emission should be expected not to escape the starbursts.

A possible scenario for the origin of the UV and FIR emissions could be a patchy dust distribution causing some of the UV to be completely extincted and some to leak relatively un-extincted, in a similar way as proposed by the holes in the dust cover by \citet{2017ApJ...847...21F}.

The UV/FIR lack of spatial coincidence has important implications for energy balance codes as noted by \citet{2016ApJ...833..103H}, where the detected stellar light will have no information about the obscured starburst \citep{2015ApJ...799...81S}.

Therefore, the results here support that IRX and $\beta$ are unrelated for such FIR-bright sources and that extinction correction prescriptions based on the nominal IRX-$\beta$ relation are inappropiate for DSFGs.

In Section~\ref{subsec:betamap} we interpreted the UV slope differences over the source extent as variations in the dust content not detected in emission in the ALMA observations. It is possible that this regime of star formation is compatible with the M99 relation. In order to check this, we calculated the expected $L_{\rm{IR}}$ below the 3$\sigma$ dust continuum detection limit over the components detected in the rest-frame UV for each source, by rescaling their FIR SEDs \citep{smolcic15,2014ApJ...782...68T}. The resulting upper limits lie above the M99 relation for all cases, not being useful on putting constraints about whether these galaxies follow M99 or lie above or below it, a subject of main focus in current studies \citep[e.g.,][]{2015Natur.522..455C,2017ApJ...845...41B,2017ApJ...847...21F,2017MNRAS.472..483F}.

\section{Stellar Masses and Merger Ratios} \label{sec:mstar}

\subsection{What Triggers $z > 4$ Starbursts?} \label{subsec:mstar_trigger}

Major mergers between gas-rich galaxies are often assumed to be the triggering mechanism for starburst galaxies, as local Universe infrared-luminous galaxies are exclusively associated with major mergers with $L_{\rm{IR}} > 10^{11.5}$\,$L_{\odot}$ \citep[e.g.,][]{1996ARA&A..34..749S}. The multiplicity of close, approximately equally-bright galaxies in the \textit{HST} images studied here would naively support a similar triggering mechanism at $z > 4$. However, as the images trace the rest-frame UV, a stellar mass analysis of the individual merging components is needed to test this picture.

In Table~\ref{tab:prop} we list the stellar masses of the stellar components of each system derived from the SED fits described in Section~\ref{subsec:sed_fit}. Also listed is the stellar mass ratio relative to the most massive component in the system ($M_{\rm{*,prim}}$).

The median stellar mass of the most massive component is $\log (M_{*}/M_{\odot}) = 10.49 \pm 0.32$ (where the uncertainty is the median absolute deviation). For the remaining less massive components the median is $\log (M_{*}/M_{\odot}) = 9.56 \pm 0.10$. A stellar mass ratio of 1:3--4 is often adopted to distinguish between major and minor mergers \citep[e.g.,][]{2003AJ....126.1183C,2008ApJ...680..246T,2009MNRAS.394.1713K,2011ApJ...742..103L,2016ApJ...830...89M}. Adopting this definition, AzTEC5 is formally classified as major merger, with a stellar mass ratio for the two most massive components (AzTEC5-1 and AzTEC5-2) of $M_{\rm{*}}/M_{\rm{*,prim}} = 3.0$. Vd-17871 could be classified as a major or minor merger depending on the exact distinction ratio ($M_{\rm{*}}/M_{\rm{*,prim}} = 3.5$). The rest of the systems are consistent with undergoing at least one minor merger (also including AzTEC5 which might undergo minor merging with AzTEC5-3 and AzTEC5-4). Furthermore, it is important to note that regardless of the precise distinction between major and minor mergers, the components detected in dust continuum with ALMA are undergoing starbursts with SFRs that overwhelm those of the companions, and therefore, the stellar mass ratios are expected to decrease. Taking this into account all systems could be classified as minor mergers (except AzTEC5-1 and AzTEC5-2, both starbursting systems).

In addition to the components which were bright enough to estimate stellar masses, AK03, AzTEC5 and J1000+0234 present additional low S/N companions detected in one or more of the \textit{HST} images (marked with arrows in Figure~\ref{fig:hstalma_multi}), which may be additional minor merger components if they are at the same redshift. The residuals in the modeling of the \textit{Spitzer}/IRAC images do not show significant detections at their positions, and thus, they must be less massive than the detected companions. In fact, the \textit{HST} images display $2 < \rm{S/N} < 3$ potential additional low-mass components in the case of $F814W$ particularly, as expected if they are small, blue star-forming galaxies. If their redshifts are confirmed, it would be further evidence for the starbursts in $z \sim 4.5$ SMGs being triggered by multiple minor mergers. A picture consistent with living in overdense environments \citep[e.g.,][]{2004ApJ...611..725B,smolcic17}. Indeed \citet{smolcic17} showed evidence that AzTEC1, AzTEC5, J1000+0234 and Vd-17871 have statistically significant small-scale overdensities.

Note, however, that these results do not rule out that major mergers played a role in triggering these starbursts, if they have already coalesced, or if they are so close that they are not resolved in the \textit{HST} and ALMA data. Indeed the multiple FIR peaks in AK03 and AzTEC5, and the color gradients observed in the most massive components of the systems (most prominently in AzTEC1, J1000+0234 and Vd-17871), are consistent with such a picture.

\subsection{Comparison to Previous Stellar Mass Estimates} \label{subsec:mstar_prev}

Previous estimates of the stellar mass of the galaxies in this sample, derived using MAGPHYS \citep{2008MNRAS.388.1595D}, led to a median value of $\log (M_{*}/M_{\odot}) = 10.92 \pm 0.13$ \citep{smolcic15,2014ApJ...782...68T}. This is $\sim 0.4$\,dex higher than our derived median value for the most massive component. Adding up all the components per source the median total stellar mass would be slightly higher $\log (M_{*}/M_{\odot}) = 10.63 \pm 0.11$, but still $\sim 0.3$\,dex lower than the previous estimates for this sample. Recent results from \citet{2017A&A...606A..17M}, also employing MAGPHYS, are also systematically higher by at least 0.3\,dex for the sources in common with our sample (AzTEC1, AzTEC5 and J1000+0234). Such systematic discrepancies are consistent with the expected overestimation of MAGPHYS-derived stellar masses and slight underestimate of exponentially declining models employed here, according to \citet{2014A&A...571A..75M} SMGs stellar masses studies from simulated datasets.

We also compared our stellar mass estimates with those listed in the 3D-\textit{HST} survey catalog \citep{2016ApJS..225...27M,2012ApJS..200...13B,2014ApJS..214...24S} for the sources covered in the CANDELS fields (e.g., AK03 and AzTEC5) and the COSMOS2015 catalog \citep{2016ApJS..224...24L}. In general the catalogs succesfully extract the majority of the components for these complex objects and lists photometric redshifts consistent with the available spectroscopic redshifts (see Table~\ref{tab:prop}). However, for a subset we found significant discrepancies in the derived stellar masses. The discrepancy might be due to the different approach in the photometry measurements. While we measured fluxes in apertures carefully chosen to minimize the effect of blending and applied aperture corrections, COSMOS2015 employs automated PSF-matched photometry, which can be more contaminated by blending of close objects.

Furthermore, J1000+0234-N is not in the COSMOS2015 catalog, and the bulk of its stellar mass is associated to J1000+0234-S (likely due to a mismatch between the \textit{Spitzer}/IRAC and optical/near-IR data). AzTEC/C159 is also missing from the catalog, due to its extreme faintness in the optical/near-IR. Similarly, there is no entry corresponding to the location of AzTEC5-1 in either 3D-\textit{HST} or COSMOS2015. The absence and mis-identifications of massive and optically faint sources could affect the photometry, and thus, the stellar mass estimates. It could also affect the stellar mass functions at high redshifts \citep[e.g.,][]{2017A&A...605A..70D}.

J1000+0234 is also present in the recent work by \citet{2017A&A...608A..15B} and the assigned shorter wavelength counterpart to the ALMA detection is also J1000+0234-S, since J1000+0234-N remains undetected. This indicates that significant offsets between sub-mm/radio sources and UV/optical/near-IR counterparts could be indeed due to the presence of multiple blended, and perhaps merging, components if the depth and resolution of the data are not enough to detect all those components (provided a good relative astrometry between the different instruments).

Compared with previous estimates of the average stellar masses of SMGs, our results are in line with studies indicating that most SMGs have $M_{*} < 10^{11}$\,$M_{\odot}$ \citep[e.g.,][]{2011MNRAS.415.1479W,2011ApJ...740...96H,2013MNRAS.436.1919C,2014ApJ...788..125S}. Other studies report higher values $M_{*} > 10^{11}$\,$M_{\odot}$ also for $z \sim 4.5$ sources \citep{2010A&A...514A..67M,2012A&A...541A..85M,2017MNRAS.469..492M}. The median stellar mass of the satellite galaxies is consistent with estimates for faint LBGs at similar redshifts \citep{2010MNRAS.401.1521M}.

\subsection{Stellar Mass Uncertainties and Caveats} \label{subsec:mstar_cav}

Stellar masses of highly obscured starburst galaxies are notoriously difficult to estimate. In this work we took advantage of high resolution \textit{HST} imaging to identify the positions of multiple stellar components in the systems, which in turn was used to deblend the rest-frame optical \textit{Spitzer}/IRAC fluxes that are tracing the stellar mass available for these high-redshift systems. However, our stellar mass estimates are potentially subject to a number of additional systematic uncertainties.

One caveat is that some of the components lack spectroscopic confirmation. That is the case of AK03-S, AzTEC1-N and all components of AzTEC5. When possible we assumed that these components were at the same redshift as their spectroscopically confirmed companions. For AK03-S \citet{smolcic15} found a $z_{\rm{phot}} = 4.40 \pm 0.10$ or $z_{\rm{phot}} = 4.65 \pm 0.10$, depending on the template used. Therefore, the two components are likely at the same redshift. AzTEC1-N is a very faint component with $\rm{S/N} < 3$ in all the \textit{HST} bands, but it is detected above this threshold in HSC $r$, $i$ and $z$ bands, and in the IRAC bands, where the residuals from AzTEC1-S fitting showed that there is indeed a secondary component towards the North. We derived a photometric redshift consistent with being at the same redshift than AzTEC1-S within the uncertainties. Its probability distribution peaks at 3.77$_{-0.22}^{+0.32}$ (where the uncertainties are the 1$\sigma$ percentiles of the maximum likelihood distribution), being not null in the redshift range $3.2 < z < 4.7$). In the case of AzTEC5 none of the components have spectroscopic redshifts, but the 3D-\textit{HST} survey catalog \citep{2016ApJS..225...27M,2012ApJS..200...13B,2014ApJS..214...24S} lists $z_{\rm{phot}} = 3.63_{-0.15}^{+0.14}$ for AzTEC5-2, $z_{\rm{phot}} = 4.02_{-0.08}^{+0.08}$ for AzTEC5-3 and $z_{\rm{phot}} = 3.66_{-0.43}^{+0.40}$ for AzTEC5-4. Therefore, it seems plausible that all components in AzTEC5 lie at the same redshift within the uncertainties.

Another caveat in the stellar mass estimates come from the assumptions made in the SED fits. \citet{2014A&A...571A..75M} studied the importance of the assumed SFHs \citep[see also][]{2011ApJ...740...96H} over several SED fitting codes, concluding that the exponentially declining SFHs used here are able to recover the stellar masses of their simulated SMGs, with slight underestimation and significant scatter. Regardless of the model employed, the derived photometry and the color of the sources already indicates that there is a component more massive that the other. The most massive components have higher IRAC fluxes and they are also redder than their fainter IRAC companions.

Given the extreme dust mass surface densities derived for this sample (see Table~\ref{tab:fir_size}), if the stars formed in-situ in the starburst that created the dust it is possible that some stellar mass is so obscured that it is not detectable even by IRAC, and thus, not accounted for in the SED fitting. Higher spatial resolution rest-frame FIR continuum observations would be needed to disentangle the underlying structure of the dust emitting region and measure its degree of homogeneity or clumpiness. This could reveal how much of the stellar light is completely obscured beneath the dust and the implied systematic error in the derived stellar masses. To estimate how big this effect could be, using the empirical dust-to-stellar-mass ratio (DTS) for local ULIRGs in \citet{2017MNRAS.465...54C} $\log \rm{DTS} = -2.83$, the median stellar mass of this sample would increase to $\log (M_{*}/M_{\odot}) \sim 11.6$. However, assuming the ratio from simulations in \citet{2017MNRAS.471.3152P} $\log \rm{DTS} \sim -1.8$ the effect would not be that significant, increasing to $\log (M_{*}/M_{\odot}) \sim 10.9$.

\begin{figure}
\begin{center}
\includegraphics[width=\columnwidth]{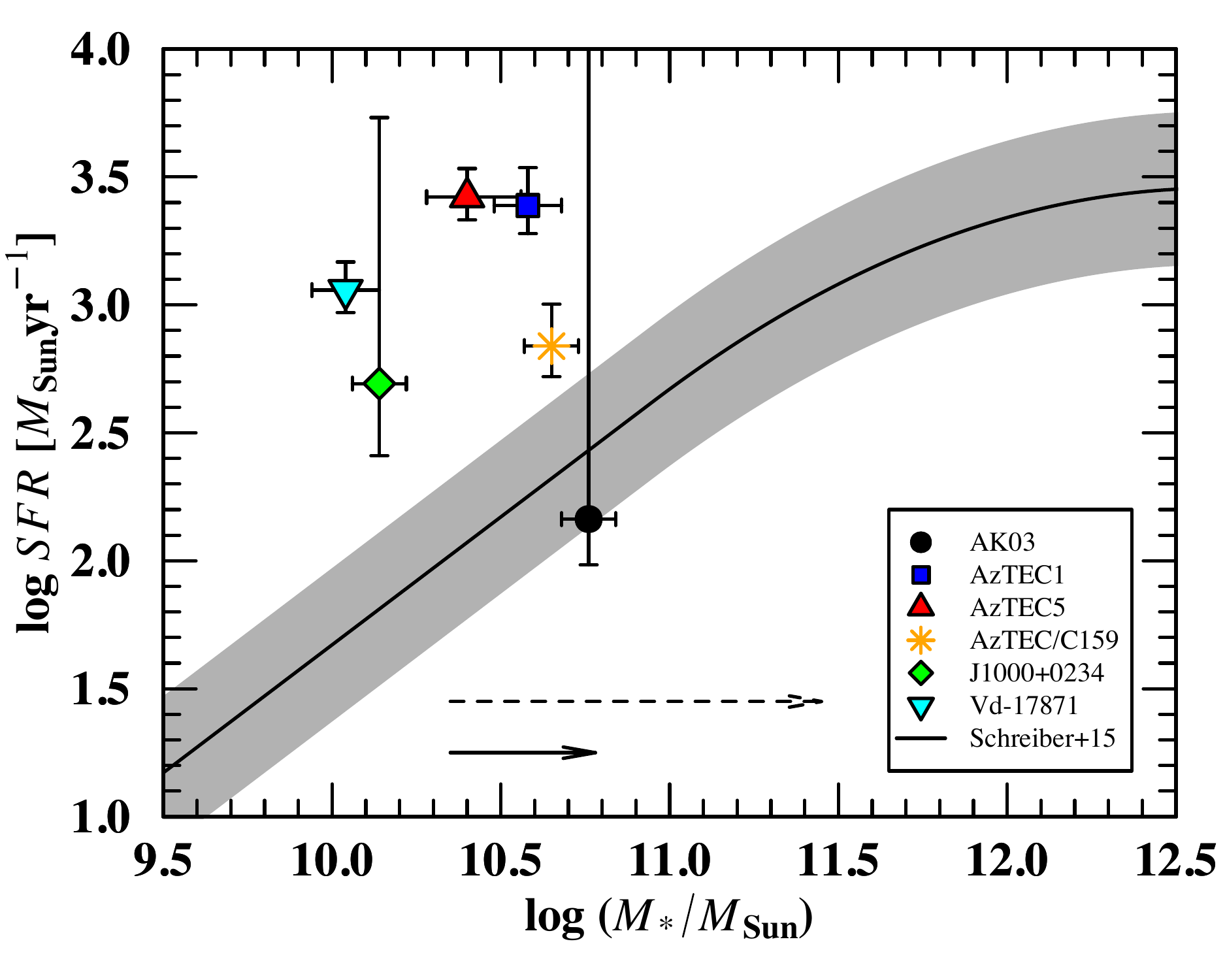}
\caption{$SFR-M_{*}$ plane. Our sample (colored symbols) lies above the main sequence of star-forming galaxies as defined by \citet{2015A&A...575A..74S} (plotted at $z = 4.5$ for reference and converted from Salpeter to Chabrier IMF). A 0.3\,dex (2 times) scatter around the MS is represented by a gray shadowed region. Bottom arrows indicate the estimated increase in the derived stellar masses if a fraction is obscured by the dust ($\log \rm{DTS} \sim -1.8$ from \citet{2017MNRAS.471.3152P}, solid arrow, and $\log \rm{DTS} = -2.83$ \citet{2017MNRAS.465...54C}, dashed arrow).}
\label{fig:ms}
\end{center}
\end{figure}

Over the last decade, several studies have uncovered a tight correlation between the SFR and the stellar mass of star-forming galaxies, the so-called main sequence (MS) of star formation \citep[e.g.,][]{2007ApJ...660L..43N,2007A&A...468...33E,2007ApJ...670..156D}. Strong outliers to the MS are present at all redshifts and this is often used as a formal definition of starburst galaxies. These systems exhibit elevated specific star formation rates ($sSFR$) compared with typical MS galaxies. For the components with ALMA detection, from the total $SFR_{\rm{IR+UV}}$ and stellar masses, we obtain $sSFR = 2.5$--100\,Gyr$^{-1}$. Considering the MS as defined in \citet{2015A&A...575A..74S}, the distance to the MS ranges $sSFR/sSFR_{\rm{MS}} = 0.5$--22, calculated at the redshift of each source. Consequently, all the sources studied here would formally fall into the starburst regime, with AK03 on the MS but also consistent with the starburst region given its large SFR uncertainty (see Figure~\ref{fig:ms}). If an important fraction of the stellar mass is undetectable hidden beneath the dust, the objects would move towards smaller distances to the MS, as represented by the bottom arrows in Figure~\ref{fig:ms}.

\section{Stellar Mass-Size Plane: Evolution to Compact Quiescent Galaxies} \label{sec:mass_size}

The similar stellar mass and rest-frame optical/UV size distribution of $z > 3$ SMGs and cQGs at $z \sim 2$ has been used to argue for a direct evolutionary connection between the two populations \citep{2014ApJ...782...68T}. However, the stellar mass builds up in the nuclear starburst. At the derived SFR and stellar mass for our sample, approximately half of the descendant stellar mass would be formed during the starburst phase. The FIR size traces the region where the starburst is taking place, and thus, it is the relevant measurement to compare to the optical size in the descendant 1--2\,Gyr later, as it is the best proxy for the location of the bulk of the stellar mass once the starburst is finished.

\begin{figure}
\begin{center}
\includegraphics[width=\columnwidth]{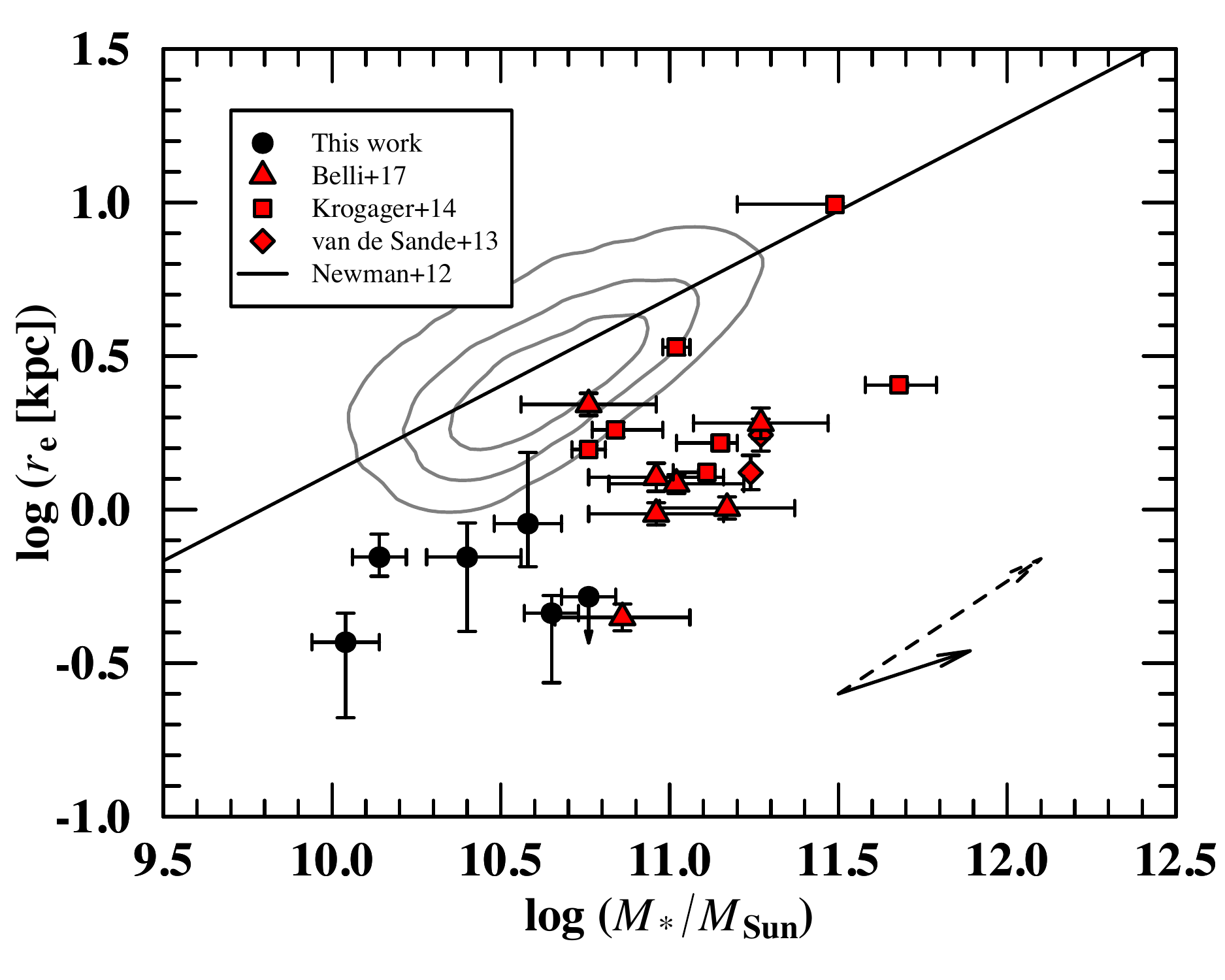}
\caption{Stellar mass-size plane location of the SMG sample in this work (black filled circles), along with $z \sim 2$ CQGs (red filled symbols) from \citet{2013ApJ...771...85V} (diamonds), \citet{2014ApJ...797...17K} (squares) and \citet{2017ApJ...834...18B} (triangles). The bottom-right black solid arrow indicates the expected evolution of the SMG sample, accounting for the stellar mass growth through the derived $SFR_{\rm{IR+UV}}$ over a duty cicle of 42\,Myr and minor merger contribution, and in size via minor mergers. Above a dashed arrow indicating the predicted evolution from a potential scenario with longer depletion time-scales of 100\,Myr and larger sizes assuming a [\ion{C}{2}] size proxy for the effective star-forming region. For comparison, the local mass-size relation from \citet{2012ApJ...746..162N} is shown as a solid line, along with SDSS local massive quiescent galaxies as gray contours \citep{2003MNRAS.343..978S}. All plotted data were converted to a concordance cosmology $[\Omega_\Lambda,\Omega_M,h]=[0.7,0.3,0.7]$ and Chabrier initial mass function (IMF) when needed.}
\label{fig:mass_size}
\end{center}
\end{figure}

In Figure~\ref{fig:mass_size} we compare the stellar masses and rest-frame FIR effective radii for our sample of SMGs to the stellar masses and rest-frame optical effective radii measured for spectroscopically confirmed cQGs at $1.8 < z < 2.5$ \citep[samples from,][]{2013ApJ...771...85V,2014ApJ...797...17K,2017ApJ...834...18B}. Note that the optical sizes in these cQGs comparison samples were also obtained by fitting the two-dimensional surface brightness distribution with \texttt{GALFIT}, as we did for the FIR sizes of our SMGs sample.

The SMGs appear offset to smaller stellar masses and sizes than cQGs, with approximately the same scatter. The median stellar mass of our SMGs is $\log (M_{*}/M_{\odot}) = 10.49 \pm 0.32$ compared to $\log (M_{*}/M_{\odot}) = 11.07 \pm 0.08$ for the cQGs. The median rest-frame FIR size for the SMGs is $r_{\rm{e}} = 0.70 \pm 0.29$\,kpc, compared to rest-frame optical sizes of $r_{\rm{e}} = 1.61 \pm 0.68$\,kpc for the cQGs. The SMGs would have to increase both in stellar mass and size to evolve into $z \sim 2$ cQGs.

In the following we discuss if such an evolution is plausible, given the observed properties of the SMG sample.

As the galaxies are undergoing starbursts, they will grow significantly in stellar mass before quenching. \citet{2014ApJ...782...68T} derived a depletion time-scale of $\tau_{\rm{gas}} = 42_{-0.29}^{+0.40}$\,Myr for the number density of $z \gtrsim 3$ SMGs and cQGs at $z \sim 2$ to match. Assuming this number, at their current median $SFR_{\rm{IR+UV}} = 920$\,$M_{\odot}$ yr$^{-1}$, the stellar mass is expected to increase by a factor of $\sim 2.24$ ($\sim 0.35$\,dex). Star formation is not expected to increase the sizes significantly. The sizes of the remnants are, however, foreseen to grow due to ongoing minor mergers.

The median stellar mass ratio of the ongoing minor mergers is 6.5 and the average number of them is 1.2. Taking into account these mergers, the expected increase in stellar mass is $\sim 2.43$ ($\sim 0.39$\,dex). Adopting the simple models of \citet{2009ApJ...697.1290B} for size growth due to minor mergers, the remnants are expected to grow by a factor of $\sim 1.39$ ($\sim 0.14$\,dex).

Simulations suggest a typical minor merger time-scale of $0.49 \pm 0.24$\,Gyr \citep{2010MNRAS.404..575L}. This provides sufficient time for the mergers to complete between $z \sim 4.5$--3.5 while not violating the stellar ages of 1--2\,Gyr derived for $z \sim 2.5$--2.0 cQGs \citep{2012ApJ...754....3T}.

The combined average stellar mass and size growth anticipated from completion of the starburst and the minor mergers is shown as the bottom-right solid arrow in Figure~\ref{fig:mass_size}. The SMGs would grow to a stellar mass of $\log (M_{*}/M_{\odot}) = 10.88 \pm 0.32$ and a size of $r_{\rm{e}} = 0.98 \pm 0.29$\,kpc, bringing the two populations into agreement within the uncertainties.

The scenario laid out here is in line with recent theoretical work by \citet{2017ApJ...839...71F}, which suggests that models with starburst-induced compaction followed by minor merger growth better reproduces the sizes of the quenched remnants than models without structural changes.

In order to provide the stellar mass increase the SMGs need enough gas reservoir to fuel the star formation. The median gas mass for our sample calculated from $M_{\rm{dust}}$ using a $\rm{GDR} = 90$ is $3.7 \times 10^{11}$\,$M_{\odot}$. The factor $\sim 2.24$ mentioned above means the creation of $3.8 \times 10^{10}$\,$M_{\odot}$, which would be achieved with a $\sim 10$\% efficiency of converting gas into stars. The available molecular gas estimates derived from $^{12}$CO measurements in the literature for our sample are: AzTEC1, $M_{\rm{H2}} = 1.4 \pm 0.2 \times 10^{11}$\,$M_{\odot}$, with $\tau_{\rm{gas}} \sim 200$\,Myr \citep{2015MNRAS.454.3485Y}; AzTEC/C159, $M_{\rm{H2}} = 1.5 \pm 0.3 \times 10^{11}$\,$M_{\odot}$, with $\tau_{\rm{gas}} = 200 \pm 100$\,Myr \citep{2017arXiv171010181J}; and J1000+0234, $M_{\rm{H2}} = 2.6 \times 10^{10}$\,$M_{\odot}$, with $\tau_{\rm{gas}} \sim 30$\,Myr \citep{2008ApJ...689L...5S}. The amount of gas available to form stars seems enough to account for the expected increase in stellar mass and the short depletion time-scale match the short duration of the SMG phase of $\sim 100$\,Myr \citep[e.g.,][]{2006ApJ...640..228T,2008ApJ...680..246T}.

In the propose scenario we assume that the rest-frame FIR dust continuum is a reasonable proxy for the effective star-forming region. [\ion{C}{2}] size estimates for a subset of our sample (Karim et al. in prep) are typically two times larger, which is in agreement with other studies finding larger [\ion{C}{2}] sizes compared with dust continuum sizes \citep[e.g.,][]{2014ApJ...796...84R,2016ApJ...816L...6D,2016ApJ...827...34O}. Considering a scenario with $\tau_{\rm{gas}} = 100$\,Myr and [\ion{C}{2}] sizes would mean a factor of $\sim 3.96$ ($\sim 0.60$\,dex) change in stellar mass and $\sim 2.78$ ($\sim 0.44$\,dex) in size, still suitable for the two populations to match, with the SMGs having a final stellar mass of $\log (M_{*}/M_{\odot}) = 11.09 \pm 0.32$ and size of $r_{\rm{e}} = 1.95 \pm 0.29$\,kpc.

\section{Discussion} \label{sec:discussion}

In this work we present detailed observations of a small sample of $z \sim 4.5$--3.5 SMGs and argue that their properties are consistent with being progenitors of $\sim 2.5$--2.0 cQGs.

We demonstrated that the distribution of the two populations in the stellar mass-size plane are consistent when accounting for stellar mass and size growth expected from the completion of the ongoing starbursts and subsequent merging with minor companions.

These conclusions are based on small samples for both the SMGs and cQGs, possibly subject to selection biases, and apply only in two broad redshift intervals. To further explore the evolutionary connection between the two populations, larger uniform samples, with a finer redshift sampling are needed. For example, cQGs are now being identified out to $z \sim 4$ \citep{2015ApJ...808L..29S}, although confirming quiescent galaxies at this high redshift can be challenging \citep{2017Natur.544...71G,2017ApJ...844L..10S,2017arXiv170903505S}. If the proposed connection holds at all redshifts, the properties of these should match those of SMGs at $z > 6$ \citep[e.g.,][]{2013Natur.496..329R,2017Natur.545..457D,2017ApJ...842L..15S,2017ApJ...850....1R}. Similarly, the properties of $z \sim2 $ SMGs should match those of 1\,Gyr old quiescent galaxies at $z \sim 1.5$.

A crucial measure placing starburst galaxies in a cosmic evolution context is their stellar mass. Unfortunately, it is a very difficult to derive due to large amounts of dust, that may prevent an unknown fraction of the stellar light to escape, even at rest-frame near-IR wavelengths. Perhaps the best way forward is to measure it indirectly, as the difference between the total dynamical mass and the gas mass (and dark matter), both of which can be estimated from molecular line observations with ALMA (Karim et al., in prep).

What triggers high-redshift starbursts remains unclear. All of the galaxies studied here showed evidence of ongoing minor mergers and this could be the process responsible of igniting the starburst, while only one showed evidence of an ongoing major merger. \citet{2017gefb.confE..15B} have recently stated that while strong starbursts are likely to occur in a major merger, they can also originate from minor mergers if more than two galaxies interact. This suggests that the triggering processes at high redshift are different from low redshift, where the most luminous starburst galaxies are almost exclusively associated with major mergers, which would also be in agreement with recent theoretical work \citep{2015Natur.525..496N}. Nevertheless, low-redshift lower luminosity LIRGs are also found to be associated with minor mergers. The difference could actually be due to the gas fraction of the most massive component in the interaction, which is higher at high redshift than at low redshift, and thus, it may allow for a relatively more intense starburst to occur in the presence of a minor merger at high redshift than at low redshift.

However, even at the relatively high spatial resolution obtained in this study, we are not able to rule out close ongoing major mergers. As an example, the nucleus of the archetypical starburst galaxy Arp\,220 breaks into two components separated by $\sim 350$\,pc \citep{2017ApJ...836...66S}. At $z = 4.5$ this corresponds to an angular separation of $\sim$ 0\farcs05, and thus, we would not be able to resolve this particular case at our current resolution (median synthesized beam size 0\farcs30$\times$0\farcs27). However, the nearby FIR peaks in two of our systems that we are able to resolve and the color gradients over all the galaxies would be consistent with such a picture.

An alternative plausible scenario would be that the starburst episode we are witnessing would be indeed triggered by previous minor or major mergers that we are currently unable to detect. The minor companions we detect here would be mergers in an early phase prior to coalescence, but not responsible for the observed starburst episode. Gas dynamics in these systems show evidence for rotationally supported star-forming disks \citep[][Karim et al., in prep]{2017ApJ...850..180J,2017arXiv171010181J}, which would have to be triggered either by gravitational instabilities or highly disipational mergers that quickly set into a disk configuration. Smooth accretion can also trigger high SFR while still maintaining a rotationally supported disk \citep[e.g.,][]{2014ApJ...790L..32R}. Some simulations of galaxy formation at high redshift have also shown that gas and stellar disks already exist at $z \gtrsim 6$ \citep[e.g.,][]{2011ApJ...731...54P,2011ApJ...738L..19R,2015ApJ...808L..17F,2017MNRAS.465.2540P}.

Recently, a population of compact star-forming galaxies (cSFGs) at $2.0 < z < 3.0$ have been suggested as progenitors for cQGs \citep[e.g.,][]{2013ApJ...765..104B,2015ApJ...813...23V,2016ApJ...827L..32B}. Two different progenitor populations are not necessarily mutually exclusive. Both SMGs and cSFGs could be part of the same global population but observed in a different phase or intensity of the stellar mass assembly, with the SMGs reflecting the peak of the process and the cSFGs being a later stage. cSFGs are consistent with an intermediate population between $z > 3$ SMGs and $z \sim 2$ cQGs, caught in a phase where the star formation is winding down and a compact remnant is emerging, transitioning from the region above the MS of star-forming galaxies \citep{2017ApJ...851L..40B} to the MS \citep{2017A&A...602A..11P}, and eventually below it. In fact \citet{2017arXiv171110047E} have recently shown that starburst galaxies exist both above and within the MS. The increased AGN fraction in cSFGs suggest that they are entering a AGN/QSO quenching phase, which could be responsible for shutting down the residual star formation, leaving behind compact stellar remnants to develop into $z \sim 2$ cQGs \citep{2013ApJ...765..104B} \citep[see also][]{1988ApJ...325...74S,2006ApJ...652..864H,2012MNRAS.421..284H,2017MNRAS.464.1380W}.

In order to further explore the evolutionary connection between SMGs, cSFGs and cQGs, larger spectroscopic samples are needed. High spatial resolution rest-frame optical/FIR observations are paramount to unveil their different subcomponents and measure accurate optical/FIR sizes, stellar masses and uncover the underlying structure of the dust. In this context \textit{JWST} observations of DSFGs at high redshift will revolutionize our understanding of galaxy mass assembly through cosmic time.

\section{Summary and Conclusions} \label{sec:summary}

A sample of six SMGs, five of which are spectroscopically confirmed to be at $z \sim 4.5$, were imaged at high spatial resolution with \textit{HST}, probing rest-frame UV stellar emission, and with ALMA, probing the rest-frame FIR dust continuum emission. We find that:

\begin{itemize}

\item The rest-frame UV emission appears irregular and more extended than the very compact rest-frame FIR emission, which exhibits a median physical size of $r_{\rm{e}} = 0.70 \pm 0.29$\,kpc.

\item The \textit{HST} images reveal that the systems are composed of multiple merging components. The dust emission pinpointing the bulk of star formation is associated with the reddest and most massive component of the merger. The companions are bluer, lower mass galaxies, with properties typical of normal star-forming galaxies at similar redshifts.

\item We find morphological evidence suggesting that the lack of spatial coincidence between the rest-frame UV and FIR emissions is the primary cause for the elevated position of DSFGs in the IRX-$\beta$ plane. This has consequences for energy balance modelling efforts, which must account for the implied high extinction.

\item A stellar mass analysis reveals that only one of the systems is undergoing a major merger. On the other hand all the systems are undergoing at least one minor merger with a median stellar mass ratio of 1:6.5. In addition, the \textit{HST} images hint the presence of additional nearby low-mass systems.

\item The stellar masses and rest-frame FIR sizes of the $z \sim 4.5$ SMGs fall on the stellar mass-rest-frame optical size relation of $z\sim 2$ cQGs, but spanning lower stellar masses and smaller sizes. To evolve into $z \sim 2$ cQGs, the SMGs must increase both in stellar mass and size. We show that the expected growth due to the ongoing starburst and minor mergers can account for such evolution.

Minor merging thus appear to play a pivotal role in the evolution of massive elliptical galaxies throughout their full cosmic history. Both for their size evolution from $z = 2$ to $z = 0$ \citep[e.g.,][]{2009ApJ...699L.178N,2012ApJ...746..162N}, but also for their formation at higher redshifts.

\end{itemize}

\acknowledgments

We thank I. Smail for his detailed comments and suggestion that help on improving this manuscript; J. M. Simpson for providing the SCUBA2 data; C. Y. Peng and G. Barro for their advice on \texttt{GALFIT}; S. Zibetti for his support with \texttt{ADAPTSMOOTH}; C. M. Casey for providing the DSFGs comparison data plotted in Figure~\ref{fig:irxbeta}; and D. Watson, J. Hjorth, I. Davidzon, H. Rhodin, K. K. Knudsen, P. Laursen, D. B. Sanders, M. P. Haynes, R. Pavesi, T. K. D. Leung and S. Mart\'in-\'Alvarez for helpful comments and suggestions. We are greatful to the anonymous referee, whose comments have been very useful to improve our work.

CGG and ST acknowledge support from the European Research Council (ERC) Consolidator Grant funding scheme (project ConTExt, grant number: 648179). AK, EJA and FB acknowledge support by the Collaborative Research Centre 956, sub-project A1, funded by the Deutsche Forschungsgemeinschaft (DFG). Support for BM was provided by the DFG priority program 1573 "The physics of the interstellar medium". DR acknowledges support from the National Science Foundation under grant number AST-1614213. VS acknowledges support from the European Union's Seventh Frame-work program under grant agreement 337595 (ERC Starting Grant, "CoSMass"). MA acknowledges partial support from FONDECYT through grant 1140099. ERD also acknowledge support by the Collaborative Research Centre 956, sub-project C4, funded by the DFG. MJM acknowledges the support of the National Science Centre, Poland, through the POLONEZ grant 2015/19/P/ST9/04010; this project has received funding from the European Union's Horizon 2020 research and innovation programme under the Marie Sk{\l}odowska-Curie grant agreement No. 665778.

Based on observations made with the NASA/ESA \textit{Hubble Space Telescope}, obtained at the Space Telescope Science Institute, which is operated by the Association of Universities for Research in Astronomy, Inc., under NASA contract NAS 5-26555. These observations are associated with program \#13294. Support for program \#13294 was provided by NASA through a grant from the Space Telescope Science Institute, which is operated by the Association of Universities for Research in Astronomy, Inc., under NASA contract NAS 5-26555.

This research made use of the following ALMA data: ADS/JAO.ALMA\#2012.1.00978.S. ALMA is a partnership of ESO (representing its member states), NSF (USA) and NINS (Japan), together with NRC (Canada) and NSC and ASIAA (Taiwan), in cooperation with the Republic of Chile. The Joint ALMA Observatory is operated by ESO, AUI/NRAO and NAOJ.

This paper employed \texttt{Astropy}, a community-developed core Python package for Astronomy \citep{2013A&A...558A..33A}; \texttt{APLpy}, an open-source plotting package for Python \citep{2012ascl.soft08017R}; CASA \citep{2007ASPC..376..127M}; \texttt{Matplotlib} \citep{Hunter:2007}; \texttt{Numpy}; \texttt{Photutils} \citep{larry_bradley_2016_164986}; \texttt{PyBDSF}; R, a language and environment for statistical computing (R Foundation for Statistical Computing, Vienna, Austria) \citep{Rcite}.

\bibliographystyle{aasjournal}{}
\bibliography{smgsz4_hstalma.bib}{}

\end{document}